\documentclass[twocolumn,showpacs,preprintnumbers,amsmath,amssymb]{revtex4}

\usepackage{graphicx}
\usepackage{dcolumn}
\usepackage{bm}

\def\fun#1#2{\lower3.6pt\vbox{\baselineskip0pt\lineskip.9pt
\ialign{$\mathsurround=0pt#1\hfil##\hfil$\crcr#2\crcr\sim\crcr}}}
\def\lap{\mathrel{\mathpalette\fun <}}
\def\gap{\mathrel{\mathpalette\fun >}}

\def\mass{{\cal M}}
\def\rg{{r_g}}
\def\msun{{\mass_\odot}}

\def\beq{\begin{equation}}
\def\eeq{\end{equation}}

\def\mh{M_{\bullet}}

\def\sbh{SBH}
\def\sbhs{SBHs}

\def\rm{r_\mathrm{m}}

\begin{document}


\title{Spin evolution of supermassive black holes and galactic nuclei}
\author{David Merritt}
\affiliation{Department of Physics and Center for Computational Relativity and Gravitation, Rochester Institute of Technology, Rochester, NY 14623}
\author{Eugene Vasiliev}
\affiliation{Department of Physics and Center for Computational Relativity and Gravitation, Rochester Institute of Technology, Rochester, NY 14623}
\affiliation{Lebedev Physical Institute, Moscow, Russia}
\date{\today}

\begin{abstract}
The spin angular momentum $\boldsymbol{S}$ of a supermassive black hole (\sbh) precesses due to torques from orbiting stars, and the stellar orbits precess due to 
dragging of inertial frames by the spinning hole.
We solve the coupled post-Newtonian equations describing the joint evolution
of $\boldsymbol{S}$ and the stellar angular momenta $\boldsymbol{L}_j, j = 1\ldots N$
in spherical, rotating nuclear star clusters.
In the absence of gravitational interactions between the stars,
two evolutionary modes are found: (1) nearly uniform precession of $\boldsymbol{S}$
about the total angular momentum vector of the system;
(2) damped precession, leading, in less than one precessional period, to alignment
of $\boldsymbol{S}$ with the angular momentum  of the rotating cluster.
Beyond a certain distance from the \sbh, the time scale 
for angular momentum changes due to gravitational encounters between the stars 
is shorter than spin-orbit precession times.
We present a model, based on the Ornstein-Uhlenbeck equation, 
for the stochastic evolution of star clusters due to gravitational encounters
and use it to evaluate the evolution of $\boldsymbol{S}$ in nuclei where
changes in the $\boldsymbol{L}_j$ are due to frame dragging close to the 
\sbh\ and to  encounters farther out.
Long-term evolution in this case is well described as uniform precession of the \sbh\ 
 about the cluster's rotational axis, with an increasingly important stochastic contribution
when \sbh\ masses are small.
Spin precessional periods are predicted to be strongly dependent on nuclear properties,
but typical values are $\sim 10^7-10^8$ yr for low-mass \sbhs\ in dense
nuclei, $\sim 10^{8}-10^{10}$ yr for \sbh\ masses $\sim 10^8\msun$, and
$\sim 10^{10}-10^{11}$ yr for the most massive \sbhs.
We compare the evolution of \sbh\ spins in stellar nuclei to the case of torquing by 
an inclined, gaseous accretion disk.
\end{abstract}

\pacs{Valid PACS appear here}
\maketitle
\section{\label{Section:Introduction} Introduction}

An accretion disk fed by gas whose angular momentum is misaligned with
that of the central supermassive black hole  (\sbh) will experience
 Lense-Thirring \cite{LenseThirring1918} precession.
Viscous torques near the \sbh\ align the gas with the \sbh\ 
equatorial plane \cite{BardeenPetterson1975};
farther out, the gas remains inclined, producing a constant torque
that causes the \sbh\ spin axis to precess.
Such precession has been invoked as an explanation for
changes in the direction of radio jets in active galaxies \cite{BBR1980,Roos1988}.
Continued accretion of gas from a misaligned plane will eventually reorient the 
\sbh, although the time required for realignment is uncertain \cite{NatarajanPringle1998}.

Accretion disks are believed to be associated with only a small fraction of \sbhs.
Here we consider the more generic, and perhaps simpler, case of a rotating
\sbh\ embedded in a nuclear cluster of stars or stellar remnants.
If  the  cluster has a net  angular momentum that is misaligned
with the \sbh\ spin, a mutual torque will be exerted between stars and \sbh,
even if the {\it spatial} distribution of the stars is precisely spherical.

In the simplest such model, the stars move independently of each other. 
Differential precession (``phase mixing'') will nevertheless cause stellar orbits 
near the \sbh\ to distribute their angular momentum vectors $\boldsymbol{L}_j$ uniformly
about the spin $\boldsymbol{S}$,  decreasing the torque that they exert on the hole.
The angular momentum associated with stars farther out can remain misaligned, 
leading to a forced precession of the \sbh, similar to what occurs
in the case of misaligned accretion disks.

By solving the coupled post-Newtonian equations describing a spinning \sbh\ 
and a rotating cluster,
we verify that such an outcome is possible, 
at least starting  from certain initial conditions.
However we find a second evolutionary mode as well,
in which differential precession causes the inner system to reach alignment 
with the total (spin plus orbital) angular momentum, 
resulting in a steady state with no subsequent precession of the hole.

Stars also interact with each other gravitationally; these 
encounters lead to changes in stellar  angular momenta, on time scales that
can be short compared with Lense-Thirring  times.
Unlike changes due to frame-dragging, evolution of the $\boldsymbol{L}_j$ due
to encounters is essentially random.
There is a region near the \sbh, the ``sphere of rotational influence,''
in which encounter times are long compared with frame-dragging times.
Within this region, stellar orbits precess uniformly, while outside of it, changes in the $\boldsymbol{L}_j$ are due primarly to encounters and are random.
The size of this sphere varies from $\sim 10^{-3}$ pc in the nuclei of galaxies
like the Milky Way, to $\sim 10^1$ pc in nuclei containing the most massive 
\sbhs. 
We develop a stochastic model for the evolution of $\boldsymbol{S}$ that includes
the effects of encounters on the  $\boldsymbol{L}_j$.
In this model, net alignment of the stellar angular momena with the \sbh\ spin is less efficient, and the \sbh\ typically continues to precess about the mean $\boldsymbol{L}$ of the stellar cluster,
although its instantaneous precession rate can vary stochastically due to the
stochastically changing $\boldsymbol{L}_j$.

Evolution of \sbh\ spins to due torquing from stars has many parallels
with evolution due to torquing from an accretion disk, surprisingly so given
that one process is energy-conserving and the other is dissipative.
We compare and contrast the two sorts of evolution in the ``Discussion'' section,
where we also summarize observational and theoretical evidence for nuclear rotation,
and discuss the implications of our results for the experimental determination
of black hole spins.

Throughout this paper we ignore the contribution of stellar captures to the evolution
of $\boldsymbol{S}$.

\section{\label{Section:SpinOrbitEquations}Spin-orbit equations}

A Kerr black hole of mass $\mh$ has gravitational radius $\rg$ given by
\beq\label{Equation:Definerg}
\rg \equiv \frac{G\mh}{c^2} \approx 4.8\times 10^{-8} \left(\frac{\mh}{10^6\msun}\right)\mathrm{pc} 
\eeq
and spin angular momentum $\boldsymbol{S}$, which we write in terms
of the dimensionless spin parameter $\boldsymbol{\chi}$ as
\beq\label{Equation:DefS}
\boldsymbol{S} = \boldsymbol{\chi}\frac{G\mh^2}{c},\ \ \ \ 
|\boldsymbol{\chi}|\le 1.
\eeq
To lowest post-Newtonian (PN) order,
the spin evolves (precesses) in response to torques from orbiting stars
according to \cite{Kidder1995}
\begin{equation}\label{Equation:SpinEvolxv}
\frac{d\boldsymbol{S}}{dt} = \frac{2G}{c^2} \sum_{j=1}^N{m_j\over r_{j}^3} 
\left(\boldsymbol{x}_{j}\times\boldsymbol{v}_j\right)\times\boldsymbol{S}
\end{equation}
where $m_j\ll\mh$ is the mass of the $j$'th star, whose instantaneous position and velocity
relative to the \sbh\   (assumed fixed at the origin) are 
$\{\boldsymbol{x}_j, \boldsymbol{v}_j\}$, and $r_j\equiv |\boldsymbol{x}_j|$.
Eq.~(\ref{Equation:SpinEvolxv}) is invariant to the choice of spin
supplementarity condition \cite{Kidder1995}.
It can be written in the equivalent form
\begin{eqnarray}\label{Equation:SpinEvol2}
\dot{\boldsymbol S} = \boldsymbol{\omega}_S\times\boldsymbol{S},\ \ \ \ 
\boldsymbol{\omega}_S = \frac{2G}{c^2} \sum_j \frac{\boldsymbol{L}_j}{r_{j}^3} 
\end{eqnarray}
where 
\beq
\boldsymbol{L}_j \equiv m_j (\boldsymbol{x}_j\times \boldsymbol{v}_j)
\eeq
is the Newtonian angular momentum of the $j$'th star.

We are mainly interested in changes that take place on time scales long compared
with stellar orbital periods, $P$, where
\beq\label{eq:period}
P = {2\pi a^{3/2}\over \sqrt{G\mh}}
\approx 2.96\ \left(\frac{a}{\mathrm{mpc}}\right)^{3/2} \left(\frac{\mh}{10^6\msun}\right)^{-1/2}\mathrm{yr}.
\eeq
Here $a$ is the orbital semimajor axis and mpc $=10^{-3}$ pc.
Accordingly, each of the $j$ terms on the right hand side of Eq.~(\ref{Equation:SpinEvol2}) can be averaged over the unperturbed (Keplerian) orbit, 
whose semimajor axis and eccentricity are $a_j$ and $e_j$.
Using
\beq
\overline{r^{-3}} = a^{-3}\left(1-e^2\right)^{-3/2} 
\eeq
and fixing $\boldsymbol{L}_j$ during the averaging, 
the spin evolution equation becomes
\begin{subequations}\label{Equation:SpinEvol}
\begin{eqnarray}
\overline{\dot{\boldsymbol S}} &=& \overline{\boldsymbol{\omega}}_S
\times\boldsymbol{S},\\
\overline{\boldsymbol{\omega}}_S &=& \frac{2G}{c^2} \sum_j \frac{\boldsymbol{L}_j}{a_{j}^3\left(1-e_j^2\right)^{3/2}} \, .
\end{eqnarray}
\end{subequations}
Henceforth averaging over the Keplerian motion 
will be understood unless otherwise indicated.

Stellar orbits also precess in response to  frame-dragging torques from
the spinning \sbh.
Working again to lowest PN order and averaging
over the unperturbed motion yields the standard
expression for the Lense-Thirring \cite{LenseThirring1918}
precession:
\begin{subequations}\label{Equation:Ljdot}
\begin{eqnarray}
\dot{\boldsymbol L}_j &=& \boldsymbol{\omega}_j\times\boldsymbol{L}_j,
\label{Equation:Ljdota} \\
{\boldsymbol \omega}_j &=& \frac{2G\boldsymbol{S}}{c^2a_j^3(1-e_j^2)^{3/2}}.
\end{eqnarray}
\end{subequations}
For fixed $\boldsymbol{S}$,
precession described by Eq. (\ref{Equation:Ljdot}) has the form of uniform
advance of the line of nodes, the latter
defined as the intersection of the orbital plane
with the equatorial plane of the \sbh.
We denote the nodal angle by $\Omega$; thus in the orbit-averaged approximation,
$\dot{\Omega_j}=\omega_j$.
Orbits also experience precession of the argument of periastron due to both
the Schwarzschild and Kerr components of the \sbh\ metric,
but such precession leaves the $\boldsymbol{L}_j$ unchanged.

In the absence of interactions between stars,
the coupled equations~(\ref{Equation:SpinEvol}), (\ref{Equation:Ljdot}) 
determine the joint evolution of the \sbh\  spin and the stellar angular momenta.
Conserved quantities include the total angular momentum of the system:
\begin{equation}\label{Equation:JConstraint}
\boldsymbol{J} =\boldsymbol{S} +\sum_j\boldsymbol{L}_j
\equiv\boldsymbol{S} + \boldsymbol{L}_\mathrm{tot}
\end{equation}
as well as
\begin{subequations}\label{Equation:OtherConstraints}
\begin{eqnarray}
\left|\boldsymbol{S}\right| &\equiv& S \\
\left|\boldsymbol{L}_j\right| &\equiv& L_j,\  j = 1,\ldots, N.
\end{eqnarray}
\end{subequations}
Neither $\boldsymbol{S}$, $\boldsymbol{L}_\mathrm{tot}$ 
nor $|L_\mathrm{tot}|$  is conserved.
However conservation of $S$ and $\boldsymbol{J}$ implies
\beq
|J-S| \le L_\mathrm{tot} \le J+S.
\eeq

Consider the case in which all stars have the same $a$ and $e$;
for instance, the orbits could lie in a circular ring.
There is no differential precession, 
and Eqs.~(\ref{Equation:SpinEvol}), (\ref{Equation:Ljdot}) can be written
\begin{equation}
\dot{\boldsymbol S} = \boldsymbol{\omega}_0\times\boldsymbol{S},\ \ \ \ 
\dot{\boldsymbol L}_\mathrm{tot} =  \boldsymbol{\omega}_0\times\boldsymbol{L}_\mathrm{tot}
\end{equation}
where
\begin{subequations}\label{Equation:OmegaK}
\begin{eqnarray}
\boldsymbol{\omega}_0 &=& \frac{\boldsymbol{J}}{S}\,\omega_\mathrm{LT},\\
\omega_\mathrm{LT} &=& \frac{2G^2\mh^2}{c^3a^3(1-e^2)^{3/2}}\, \chi \label{Equation:OmegaKb}\\
&\approx& \frac{(7.0\times 10^5 \mathrm{yr})^{-1}}
{\left(1-e^2\right)^{3/2}} \chi
\left(\frac{\mh}{10^6\msun}\right)^2
\left(\frac{a}{1\ \mathrm{mpc}}\right)^{-3}. \nonumber
\end{eqnarray}
\end{subequations}
In this special case, $L_\mathrm{tot}$ {\it is} conserved, and both $\boldsymbol{S}$ and
$\boldsymbol{L}_\mathrm{tot}$ precess with frequency $\omega_0$ about the
fixed vector $\boldsymbol{J}$. 
The controlling parameter is $\Theta\equiv L_\mathrm{tot}/S$.
If $\Theta \ll 1$, $\dot{\boldsymbol S} \approx 0$ and 
$\boldsymbol{L}_\mathrm{tot}$ precesses about the fixed \sbh\ spin vector at the
Lense-Thirring rate;
while if $\Theta \gg 1$, $\dot{\boldsymbol L}_\mathrm{tot}\approx 0$ and
$\boldsymbol{S}$ precesses about the fixed angular momentum vector of the
stars with frequency 
$\Theta\,\omega_\mathrm{LT}\gg\omega_\mathrm{LT}$.

This simple model might apply to the
``clockwise stellar disk'' at the center of the Milky Way, which has a mass $\sim 10^4\msun$, radius 
$0.04\, \mathrm{pc} \lap r \lap 0.5$ pc,
 and mean orbital
eccentricity $\sim 0.2$ \cite{Paumard2006,Lu2009,Bartko2010}.
Setting $\mh=4\times 10^6\msun$ \cite{Gillessen2009a}, the implied $\Theta$ is
\beq
\Theta_\mathrm{CWD}\approx 2\chi^{-1} \frac{M_\mathrm{CWD}}{10^4\msun}
\left(\frac{R_\mathrm{CWD}}{0.1 \mathrm{pc}}\right)^{1/2},
\eeq 
consistent within the uncertainties with unity even if $\chi$ is as large as $1$.
Evidently, the stars in this disk torque
the \sbh\ about as much as they are torqued by it.
The mutual precession time is
\beq\label{Equation:OmegaLT}
\frac{\pi}{\omega_\mathrm{LT}} \approx 8\times 10^{10}\,\mathrm{yr}\, \chi^{-1}
\left(\frac{R_\mathrm{CWD}}{0.1\ \mathrm{pc}}\right)^{3},
\eeq
much longer than the $\sim 10^{7}$ yr age of the disk inferred from the properties
of its stars, and also long compared with other physical processes that are likely
to alter the stellar orbits (as discussed in more detail below).
Nevertheless, this example demonstrates that identified structures near 
the Galactic center \sbh\ can easily contain a net orbital angular momentum that exceeds $\boldsymbol{S}$.

The distribution of stars at distances $\lap 0.1$ pc from Sgr A$^\star$ is poorly
constrained \cite{Schoedel2007,Schoedel2009,Merritt2010}, but the total stellar mass in this region is almost certainly large compared with the $\sim 10^4\msun$ associated with the clockwise disk.
Given the strong ($\propto r^{-3}$) radial dependence of the frame-dragging torques,
even a modest degree of net circulation of the stars in this region could therefore induce a precession of the \sbh\ on time scales very short compared with the time of Eq. (\ref{Equation:OmegaLT}).

We emphasize that there is no need for the torquing stars to lie in a {\it geometrically}
flattened structure: according to Eqs. (\ref{Equation:SpinEvol}), 
all that is needed is a non-random orientation of the orbital angular momentum vectors, 
which occurs even in a precisely spherical nucleus if there is a preferred sense of orbital circulation.

In general, 
different stars will have different $a$ and $e$, 
implying different rates of nodal precession.
Close enough to the \sbh, orbital precession times will be short compared with
the precessional period of the \sbh, and the orbits will tend to distribute their
angular momentum vectors uniformly about the instantaneous $\boldsymbol{S}$.
The net torque from these stars will then fall essentially to zero, and continued precession
of the \sbh\ will be driven by stars farther out.
We expect the radius separating stars in these two regions to be roughly the
radius containing a total stellar angular momentum equal to $\boldsymbol{S}$.
We estimate that radius in the following section, after first presenting observationally
motivated models for stellar nuclei.

\section{\label{Section:Models}Spherical nuclei}

Most of the distributed mass at distances $r\lap0.1$ pc from the Milky Way 
 \sbh\ is believed to be in the form of stars much older than the stars in the clockwise disk.
The spatial distribution of these stars is believed to be approximately spherical
\cite{Schoedel2011}, with at least a modest degree of circulation \cite{Trippe2008,Schoedel2009}.

A simple model for the distribution of mass near the center
of a spherical galaxy is
\beq\label{Equation:GammaLaw}
\rho(r) = \rho_0\left(\frac{r}{r_0}\right)^{-\gamma}.
\eeq
Near the \sbh\ (but not so near that relativistic corrections
are required), the gravitational potential is
\beq\label{Equation:Phiofr}
\Phi(r) = -\frac{G\mh}{r}
\eeq
and orbits can be characterized by their semimajor axes and
eccentricities, as in the previous section.
If the stellar velocity distribution is assumed to be isotropic and stationary,
and if stars are distributed along orbits uniformly with respect to mean anomaly,
the joint distribution of $a$ and $e$ that generates the density 
(\ref{Equation:GammaLaw}) is 
\beq\label{Equation:Nofae}
N(a,e)\,da\,de = N_0\, a^{2-\gamma} da\, e\, de .
\eeq
The relation between $N_0$, $\rho_0$ and $r_0$ is easily shown to be 
\beq\label{Equation:mstarN0}
m_\star N_0 = \frac{8\pi^{3/2}}{2^\gamma} \frac{\Gamma(\gamma+1)}{\Gamma(\gamma-1/2)} \rho_0r_0^\gamma,\ \ \ \ \gamma>1/2
\eeq
where $m_\star$ is the stellar mass, assumed the same for all stars.
Values of $\gamma$ less than $1/2$ are not achievable if the velocity
distribution is isotropic \cite{DEGN};
we do not consider that possibility here, and in the modelling that follows,
 $\gamma$ will be restricted to the range $1/2 <\gamma<3$.

The relations (\ref{Equation:Phiofr}) - (\ref{Equation:mstarN0})
are valid  at radii smaller than the \sbh\ influence radius
$\rm$, customarily defined as the radius enclosing a stellar mass equal to $2\mh$:
\beq\label{Equation:Definerm}
M_\star(r<\rm) = 2\mh.
\eeq

A spherical cluster will exhibit net rotation if unequal numbers of stars (at each
$a$ and $e$, say) circulate in a clockwise {\it vs.} counter-clockwise sense about some
axis.
For instance, if one-half of the orbits in a spherical cluster with initially
isotropically-distributed velocities have their 
velocity vectors reversed such that all angular momentum
vectors point toward the same half-sphere,
the total angular momentum of the ensemble will be 
$|\boldsymbol{L}_\mathrm{tot}| = \frac12 \sum |\boldsymbol{L}_j|$.
Henceforth we characterize the net rotation of a spherical cluster by the 
factor $f$, defined as the fraction of orbits that have been ``flipped'' in this way;
$0\le f\le 1/2$ and $f$ is assumed to be independent of $a$ and $e$.

Characterizing the rotation in this way is ``conservative,'' in the sense that
a geometrically flattened nuclear cluster (e.g. a disk), or a spherical cluster
consisting of only circular orbits (an ``Einstein cluster'' \cite{Einstein1939}), 
can have a larger net angular momentum for the same radial distribution of mass.

\begin{figure}[h!]
\includegraphics[width=4.5cm,angle=-90.]{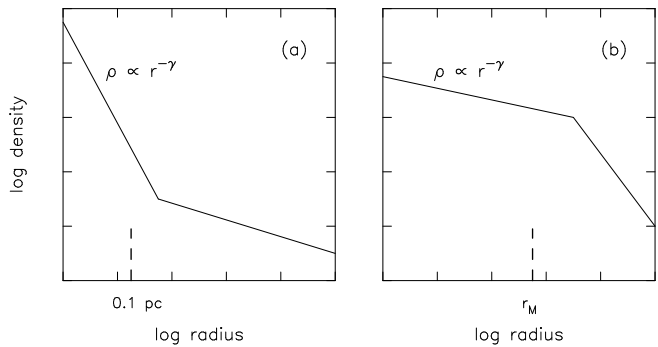}
\caption{\label{Figure:Models} 
Schematic representations of the two nuclear models considered here.
(a) Low-mass galaxy with nuclear star cluster (NSC).
 (b) High-mass galaxy with core.}
\end{figure}

Observed galaxies appear to fall into one of two classes in terms of the
parameters that define their stellar distribution at $r\lap\rm$ \cite{Graham2012}.
Massive spheroids -- elliptical galaxies, or the bulges of spiral galaxies --
with total luminosities greater than $\sim 10^{10.5}L_\odot$ have
``cores,'' regions of size $\sim\rm$ 
where the stellar density rises slowly toward the \sbh.
In these galaxies, the observed, mean relation between $\mh$ and $\rm$ is approximately
\cite{MSK2009}
\beq\label{Equation:rmvsmh}
\rm \approx 35 \left(\frac{\mh}{10^8\msun}\right)^{\alpha} \mathrm pc,
\ \ \ \ \alpha \approx 0.56
\eeq
and the index $\gamma$ that defines the central density increase varies from
$\sim 0$ at the highest luminosities to $\sim 2$ or $\sim 2.5$ at the low-luminosity
end of the range, albeit with substantial scatter \cite{Gebhardt1996,ACSFornaxII}.
The \sbhs\ in these galaxies have masses $10^{7.5}\msun\lap\mh\lap 10^{9.5}\msun$ .

Less luminous spheroids often exhibit dense central mass concentrations,
called ``nuclear star clusters'' (NSCs).
The sizes of NSCs are also comparable with $\rm$ (assuming that the
host galaxies contain \sbhs), although these structures
are too compact to be well resolved in galaxies beyond the Local Group.
The best-studied case is the Milky Way, in which the stellar density appears to
follow $\rho(r) \sim r^{-1.8}$ inside $\sim 5$ pc, compared with a \sbh\ influence radius 
of $\sim 2.5$ pc \cite{OhKimFiger2009,Schoedel2011}.
The high densities of NSCs imply short time scales for equipartition of orbital energies
 \cite{Merritt2009}, and one expects the densest NSCs to exhibit mass segregation, i.e. the heavier bodies should be more strongly concentrated toward the center than the lighter bodies.
The heaviest bodies are expected to be stellar-mass black holes (BHs),
the end products of stars with initial masses $m_\star\gap 30\msun$
whose main sequence evolution requires only a few million years;
BH masses are believed to be in the range
$5\msun \lap m_\star \lap 20\msun$ \cite{Woosley2002}, compared
with a main-sequence turnoff mass of $\sim 1 \msun$.
When energy equipartition is satisfied, the lighter population is predicted to follow
$\rho(r) \sim r^{-3/2}$ at $r\lap 0.2\rm$ while the BHs obey the steeper
relation $\rho\sim r^{-2}$ \cite{BahcallWolf1976,BahcallWolf1977}.
Detailed dynamical models of the Galactic center 
\cite{HopmanAlexander2006L, Freitag2006} suggest that if the nucleus is older
than an energy equipartition time, 
about one-half of the distributed mass inside $0.01$ pc 
would be in the form of main-sequence stars and one-half in BHs,
with a smaller mass fraction in neutron stars and white dwarves.
However it is currently unclear whether the Milky Way NSC has a relaxation time
short enough for gravitational encounters to have produced such a distribution
in 10 Gyr \cite{Merritt2010} and the distribution of observed giant stars 
(with masses $\sim 1-3\msun$) is much flatter than predicted in the relaxed models
inside $\sim 0.5$ pc \cite{Buchholz2009,Do2009,Bartko2010}.

In what follows, the central regions of bright and faint galaxies will be parametrized
in  different ways.
Nuclei of bright galaxies, with $\mh\gap 10^{7.5}\msun$, are assumed to follow
Eq.~(\ref{Equation:GammaLaw}) at $r\lap\rm$, 
 with $\rm$ determined by $\mh$ via Eq. (\ref{Equation:rmvsmh}).
The distributed mass interior to $r$ in these galaxies can be
written
\begin{eqnarray}\label{Equation:MofRCore}
M(<r) &=& 2\mh \left(\frac{r}{\rm}\right)^{3-\gamma} \\
&\approx&
2\times 10^8 \left(\frac{\mh}{10^8\msun}\right)^\beta 
\left(\frac{r}{35\,\mathrm{pc}}\right)^{3-\gamma}, \nonumber \\
\beta &=& 1 - \alpha(3-\gamma) \approx -0.68  + 0.56\gamma.\nonumber
\end{eqnarray}
Mass segregation is expected to be unimportant in 
the nuclei of giant galaxies so we set $m_\star= 1\msun$, a typical
value for an old stellar population.

In the case of galaxies with $\mh\lap 10^{7.5}\msun$, the distribution of mass
at $r<\rm$ is less certain.
We parametrize these nuclei in terms of both
$\mh$ and $M_{0.1}$, the latter defined
as the mass in stars or stellar remnants inside $r= 0.1$ pc.
If the power-law dependence of density on radius in these galaxies
were to extend outward as far as $\rm$, and if $\rm$ varied with $\mh$
as in bright galaxies,
then
\begin{subequations}\label{Equation:M0}
\begin{eqnarray}\label{Equation:M0a}
M_{0.1} &=& 2\mh \left(\frac{\rm}{0.1 \mathrm{pc}}\right)^{\gamma-3} \\
&\approx& 2\times 10^{3+\gamma} \msun \left(\frac{\mh}{10^6\msun}\right)^{1-\alpha(3-\gamma)}.
\end{eqnarray}
\end{subequations}
Eq.~(\ref{Equation:M0}b) could be taken as a rough guide to the expected value of 
$M_{0.1}$, but both $M_{0.1}$ and $\gamma$ will be considered free parameters.
We expect $1\lap\gamma\le 2$ for these nuclei; the stellar mass will be
set either to $1\msun$ (stars) or $10\msun$ (stellar BHs).

In both kinds of nuclei, rotation will be parametrized in terms of the
fraction of flipped orbits, $f$, defined above.

\begin{figure}[h!]
\includegraphics[width=4.3cm,angle=-90.]{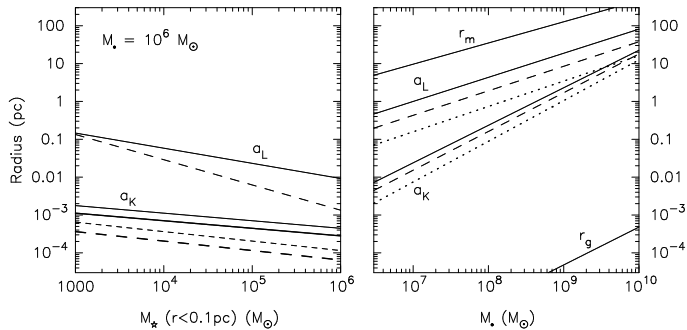}
\caption{\label{Figure:Radii} 
Characteristic radii as a function of nuclear parameters.
$a_\mathrm{L}$, Eqs. (\ref{Equation:acritlow})-(\ref{Equation:acrithigh}),
 is the semimajor axis containing a total angular momentum
equal to $S$, computed assuming $f=1/2$ and $\chi=1$ 
(maximum rotation of \sbh\ and stellar cluster).
$a_\mathrm{K}$, Eqs. (\ref{Equation:DefineaKLowb}) -
(\ref{Equation:DefineaKHighb}), 
 is the radius of rotational influence of the \sbh, assuming $\chi=1$.
The left panel assumes $\mh=10^6\msun$ and $\rg=4.8\times 10^{-8}$ pc; 
the stellar density is parametrized
in terms of $M_{0.1}$, the mass within $0.1$ pc, and $\gamma$, the power-law
index; solid lines are for $\gamma=1$ and dashed lines for $\gamma=2$.
In the case of $a_\mathrm{K}$, two values are assumed for the stellar mass: $m_\star=1\msun$ (thin lines) and $m_\star=10\msun$
(thick lines).
The right panel, for massive galaxies, assumes the relation
 (\ref{Equation:rmvsmh}) between $\mh$ and the influence radius $\rm$;
solid, dashed and dotted lines are for $\gamma=5/8$, $1$ and $3/2$ respectively.
The curves for $a_\mathrm{K}$ in the right panel assume $m_\star = 1\msun$.
The radius of tidal disruption of a solar-type star falls below the lower boundary of
both panels.}
\end{figure}

The total angular momentum associated with stars whose semimajor axes are
less than $a$ is
\begin{subequations}\label{Equation:LtotIntegral}
\begin{eqnarray}\label{Equation:LtotNofae}
\boldsymbol{L}_\mathrm{tot}(a) &=& (f\boldsymbol{e}_L)\sum_{a_j\le a} 
m_j\left[G\mh a_j(1-e_j^2)\right]^{1/2} \label{Equation:LtotIntegrala} \\ 
&\rightarrow&  (f\boldsymbol{e}_L) \left(G\mh\right)^{1/2}
N_0m_\star\int_0^a da\,a^{5/2-\gamma}\nonumber \\
&\times& \int_0^1 de\, e \left(1-e^2\right)^{1/2}  \label{Equation:LtotIntegralb} \\
&=&  \frac{4}{3(7-2\gamma)}(f\boldsymbol{e}_L) \left(G\mh\right)^{1/2}  N_0 m_\star a^{7/2-\gamma} \nonumber\\
\label{Equation:LtotIntegralc}
\end{eqnarray}
\end{subequations}
with $\boldsymbol{e}_L$ a unit vector in the
direction of $\boldsymbol{L}_\mathrm{tot}$.
We define $a_\mathrm{L}$ such that 
\beq
L_\mathrm{tot}(a_\mathrm{L}) = S = \chi\frac{ G\mh^2}{c}.
\eeq
For low-luminosity galaxies, we find
\begin{eqnarray}\label{Equation:acritlow}
&&\left(\frac{a_\mathrm{L}}{0.1\,\mathrm{pc}}\right)^{7/2-\gamma}  \approx 
1.5\times 10^{-2} \frac{\chi}{f} 
\frac{2^\gamma(7-2\gamma)}{3-\gamma}
\frac{\Gamma(\gamma-1/2)}{\Gamma(\gamma+1)}  \nonumber \\
&&\times  \left(\frac{M_{0.1}}{10^4\msun}\right)^{-1}
\left(\frac{\mh}{10^6\msun}\right)^{3/2}
\end{eqnarray}
while for bright galaxies,
\begin{eqnarray}\label{Equation:acrithigh}
&&\left(\frac{a_\mathrm{L}}{35\,\mathrm{pc}}\right)^{7/2-\gamma}  \approx 
3.9\times 10^{-5} \frac{\chi}{f} 
\frac{2^\gamma(7-2\gamma)}{3-\gamma}
\frac{\Gamma(\gamma-1/2)}{\Gamma(\gamma+1)}  \nonumber \\
&&\times  
\left(\frac{\mh}{10^8\msun}\right)^{1/2+\alpha(3-\gamma)} .
\end{eqnarray}
Figure~\ref{Figure:Radii} plots $a_\mathrm{L}$ as a function of nuclear parameters.
In massive galaxies, and for $\chi/f\approx 1$,
\beq
10^{-2}\rm\lap a_\mathrm{L} \lap 10^{-1} \rm . \nonumber
\eeq

The approximate radius of tidal disruption of a Solar-type star is
\cite{DEGN}
\beq
r_\mathrm{t} \approx 9.8\times 10^{-3} \left(\frac{\mh}{10^8\msun}\right)^{1/3} \mathrm{mpc}
\eeq
that is
\beq
\frac{r_\mathrm{t}}{\rg} \approx 2.0 \left(\frac{\mh}{10^8\msun}\right)^{-2/3}.
\nonumber
\eeq
This radius is small compared with all radii relevant to the spin evolution of
\sbhs.
Compact remnants would not be affected by tides from the \sbh\ at any radius 
greater than $\rg$.

Based on the arguments in the preceding section, we expect stars at
$r\lap a_\mathrm{L}$ to precess about the \sbh\ in a time short compared with the
precession time of the \sbh.

\section{\label{Section:SpinOrbitEvolution} Spin-orbit evolution}

The focus in this section is on the large-$N$, or ``collisionless,'' limit,
 appropriate for giant galaxies in which the central density is low and
time scales for  gravitational interactions between stars are long.
(A more precise criterion is given in \S\ref{Section:Encounters}.)
Accordingly, the number of stars in the numerical integrations was 
chosen to be large enough, 
typically $N=10^6$,  that discreteness effects were small; otherwise the value
of $N$ is unimportant.

Assuming a density law (\ref{Equation:GammaLaw}),
the coupled evolution equations
(\ref{Equation:SpinEvol}) and~(\ref{Equation:Ljdot}) admit of straightforward
scaling relations.
If the distributions of orbital
eccentricities and inclinations are invariant under the rescaling,
we can write
\begin{subequations}
\begin{eqnarray}
\omega_\mathrm{S} &\propto& \mh^{1/2} \rho_0 r_0^{\gamma}
\int a^{-(\gamma+1/2)} da, \\
\omega_j &\propto& \mh^2 a_j^{-3} \chi.
\end{eqnarray}
\end{subequations}
Consider first the case $\rm\propto \mh^\alpha$, $\alpha\approx 0.56$
that was adopted for luminous galaxies.
Setting $r_0=\rm$ in Eq. (\ref{Equation:GammaLaw})
gives $\rho_0 =\rho(\rm) \propto \mh \rm^{-3}
\propto \mh^{1-3\alpha}$.
Then
\begin{subequations}
\begin{eqnarray}
\omega_\mathrm{S} &\propto& \mh^{(3-5\alpha)/2}, \\
\omega_j &\propto& \mh^{2-3\alpha} \chi.
\end{eqnarray}
\end{subequations}
Scaling $\mh$ and $\chi$ independently as
\beq
\mh \rightarrow C_1\mh, \ \ \ \ \chi \rightarrow C_2\chi
\eeq
then yields
\begin{subequations}
\begin{eqnarray}
\omega_\mathrm{S} &\propto& C_1^{(3-5\alpha)/2}, \\
\omega_j &\propto& C_1^{2-3\alpha}C_2.
\end{eqnarray}
\end{subequations}
Evidently we require
\beq
C_2 = C_1^{\frac12(\alpha-1)} \sim C_1^{-0.2}
\eeq
if the unit of time, $[T]$, is to scale the same way in both evolution equations.
With this choice,
\beq
[T] \propto C_1^{(5\alpha - 3)/2} \sim C_1^0
\eeq
since $\alpha\approx 0.56 \approx 3/5$.

In the case of low-luminosity galaxies, the nuclear density was specified
by the independent parameter $M_{0.1}$, the stellar mass inside $0.1$ pc.
Defining a third scale factor as
\beq
M_{0.1} \rightarrow C_3 M_{0.1},
\eeq
it is clear that
\begin{subequations}
\begin{eqnarray}
\omega_\mathrm{S} &\propto& C_1^{1/2} C_3^{-1}, \\
\omega_j &\propto& C_1^{2}C_2
\end{eqnarray}
\end{subequations}
and a common unit of time requires
\beq
C_2 = C_1^{-3/2} C_3^{-1}.
\eeq
For both sorts of rescaling, the  condition $\chi < 1$ implies limits
on the values of $C_1$ and $C_3$.

Integrations of the coupled equations~(\ref{Equation:SpinEvol}), (\ref{Equation:Ljdot}) 
were carried out using a 4(5) order Runge-Kunge routine with adaptive time steps
\cite{Brankin1991}.
Monte-Carlo initial conditions for $N$ stars were first 
generated from Eq. (\ref{Equation:Nofae}) assuming
a random distribution of orbital planes, i.e. an isotropic velocity distribution.
An upper limit, $a_\mathrm{max}$, was imposed on $a$, and
a lower limit, $r_{p,\mathrm{min}}$, on the radius of orbital periapsis
$r_p=a(1-e)$.
A fraction $f$ of the orbits at each ($a,e$) were then ``flipped'' 
(the sign of $\boldsymbol{L}_j$ was changed) in order
to give the cluster a net rotation about the $z$-axis.

For these initial models,
the spin precession vector $\boldsymbol{\omega}_S$,
Eq.~(\ref{Equation:SpinEvol}b),  is given by
\begin{subequations}\label{Equation:OmegaS}
\begin{eqnarray}
\boldsymbol{\omega}_S &=&
\frac{2G}{c^2}(f\boldsymbol{e}_L)\sum_j 
\frac{m_j\left[G\mh a_j(1-e_j^2)\right]^{1/2}}{a_j^3(1-e_j^2)^{3/2}}
\\ \label{Equation:OmegaSphereb}&\rightarrow& \frac{2G^{3/2}\mh^{1/2}}{c^2} (f\boldsymbol{e}_L)
N_0m_\star\int\int \frac{da\, ede}{a^{1/2+\gamma}(1-e^2)} \nonumber \\
\end{eqnarray}
\end{subequations}
where $m_\star$ is the mass of one star and
$\boldsymbol{e}_L$ is a unit vector in the direction of $\boldsymbol{L}_\mathrm{tot}$.
The integral (\ref{Equation:OmegaSphereb}) diverges as the integration limit in
$a$ tends to zero for $\gamma\ge 1/2$, or as the limit in $e$ tends to one.
A lower limit could be placed on $a(1-e)$ by the requirement that stars come
only so close to the \sbh\ before being captured or tidally disrupted.
 But as noted above,
one expects the net angular momentum of stars at small radii to align quickly
(on a time scale much shorter than the time for changes in $\boldsymbol{S}$)
with $\boldsymbol{S}$, reducing their contribution to $d\boldsymbol{S}/dt$.

\begin{figure}
\includegraphics[width=6.cm,angle=-90.]{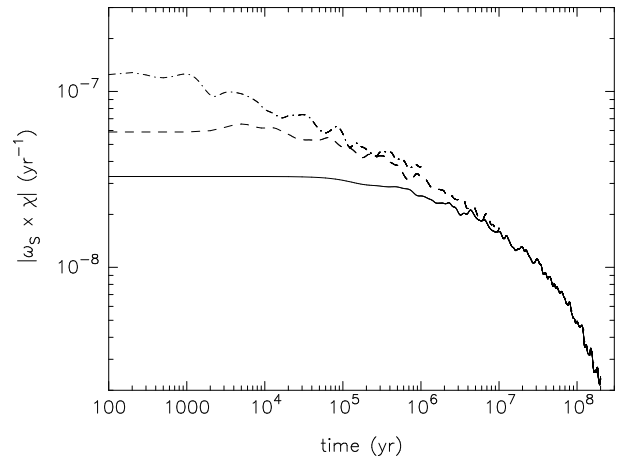}
\caption{\label{Figure:rpsdot} 
Evolution of the \sbh\ spin precession rate in a set of integrations with $r_{p,\mathrm{min}} = (0.03,0.1,0.3) $ mpc and $a_\mathrm{max}=30$ mpc.
The other model parameters are specified in the text.}
\end{figure}

That this does indeed occur is illustrated in Figure~\ref{Figure:rpsdot}, which shows integrations of
a set of models that differ only in the choice of $r_{p,\mathrm{min}}$.
The models have $\mh=10^6\msun$, $\chi=1$, $a_\mathrm{max} = 30$ mpc,
$f=1/2$, $\gamma=1$, and $N=10^6$.
The \sbh\ spin axis was oriented initially at an angle of $60^\circ$ with respect
to $\boldsymbol{L}_\mathrm{tot}$.
For these parameters, $a_\mathrm{L} \approx 5$ mpc and 
$\omega_\mathrm{LT}(a_\mathrm{L}) 
\approx 1\times 10^{-8}$ yr$^{-1}$.
The initial conditions with smaller $r_{p,\mathrm{min}}$ have larger initial
$\omega_\mathrm{S}$.
However the torque from the inner stars decays on a time scale of
order the Lense-Thirring time for the innermost orbits as their
angular momentum vectors distribute themselves uniformly about 
$\boldsymbol{S}$, and $\boldsymbol{S}$ hardly changes in this time.

\begin{figure*}
\includegraphics[width=10.cm,angle=-90.]{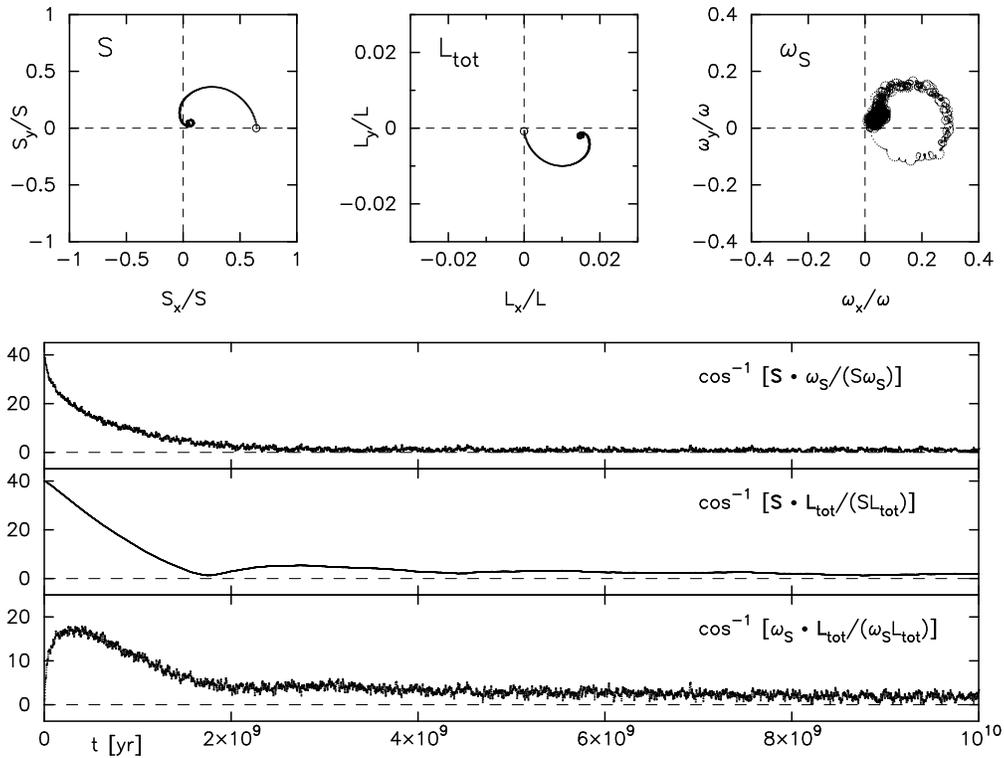}
\caption{\label{Figure:sphere40} 
Evolution of $\boldsymbol{S}$, $\boldsymbol{L}_\mathrm{tot}$,
and $\boldsymbol{\omega}_\mathrm{S}$ in a cluster where the initial
\sbh\ spin axis was offset by $\theta_0=40^\circ$ from the stellar angular momentum
vector. The other parameters of the model are given in the text.
In the upper panels, the open/filled circles indicate initial/final times respectively.
This is an example of damped precession: the vectors $\boldsymbol{S}$, $\boldsymbol{L}_\mathrm{tot}$ and $\boldsymbol{\omega}_S$ reach
a common orientation after roughly one precession cycle. 
Qualitatively the same sort
of evolution occurs for $0\le\theta_0\lesssim 45^\circ$.
}
\end{figure*}
\begin{figure*}
\includegraphics[width=10.cm,angle=-90.]{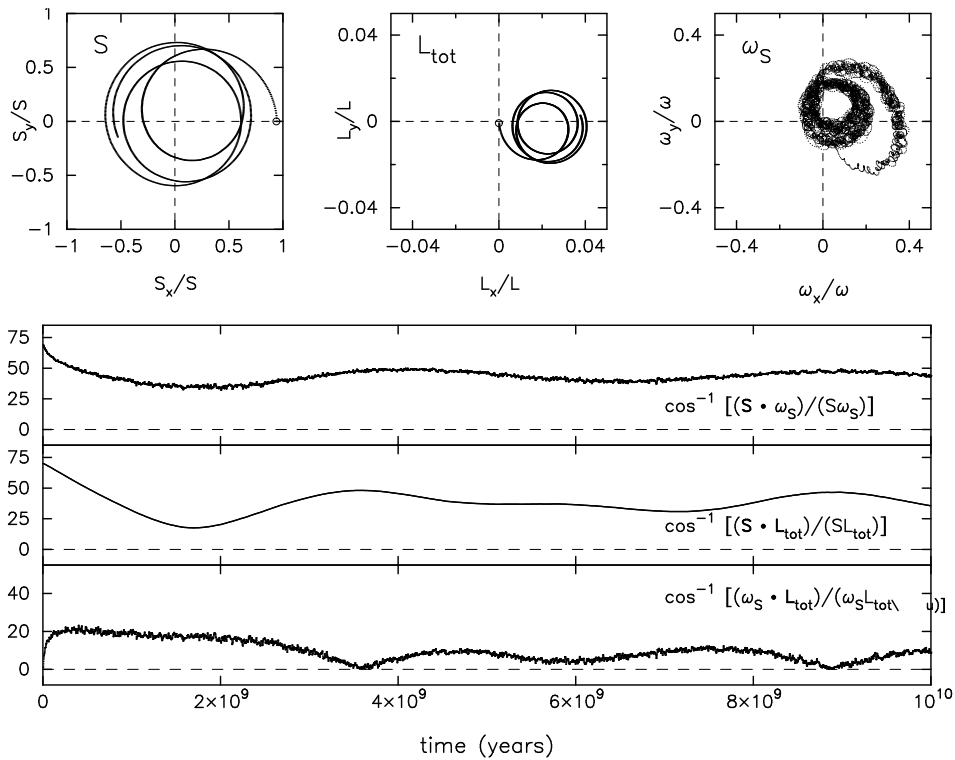}
\caption{\label{Figure:sphere70} 
Like Figure~\ref{Figure:sphere40}, except that $\theta_0=70^\circ$. In this case the
\sbh\ continues to precess about $\boldsymbol{J}$.
This mode occurs for $45^\circ\lesssim\theta_0<180^\circ$.
}
\end{figure*}

The long-term evolution of the models in Figure~\ref{Figure:rpsdot} consists of
precession of the \sbh\ about $\boldsymbol{J}\approx \boldsymbol{L}_\mathrm{tot}$.
It turns out that a second evolutionary mode is possible in spherical models like these.
This is illustrated in Figures~\ref{Figure:sphere40} and \ref{Figure:sphere70},
based on a cluster with  parameters
$\mh = 10^6\msun$, $\chi=1$, $\gamma=1$,  $f = 1/2$, $a_\mathrm{max}=100$ 
mpc, $r_{p,\mathrm{min}}=1$ mpc, and a total stellar mass of $10^5\msun$.
For this model, 
$M_{0.1} \approx 6\times 10^4\msun$ and
$a_\mathrm{L}\approx 15$ mpc.
The integrations shown in Figures~\ref{Figure:sphere40} and \ref{Figure:sphere70} 
are from a sequence in which the
initial angle, $\theta_0$, between $\boldsymbol{S}$ and $\boldsymbol{L}_\mathrm{tot}$
was varied in steps of $10^\circ$, from $10^\circ$ to $170^\circ$.
For $\theta_0\gap 45^\circ$, evolution at late times consists
of nearly uniform precession of the \sbh\ spin axis about $\boldsymbol{J}$,
as in the integrations of Figure~\ref{Figure:rpsdot}.
However if $\theta_0\lap 45^\circ$,  precession continues only for a 
single cycle or less, after which the vectors $\boldsymbol{S}, \boldsymbol{L}_\mathrm{tot}$
and $\boldsymbol{\omega}_S$ are nearly aligned and precession essentially stops.

\begin{figure}[h!]
\includegraphics[width=8.5cm]{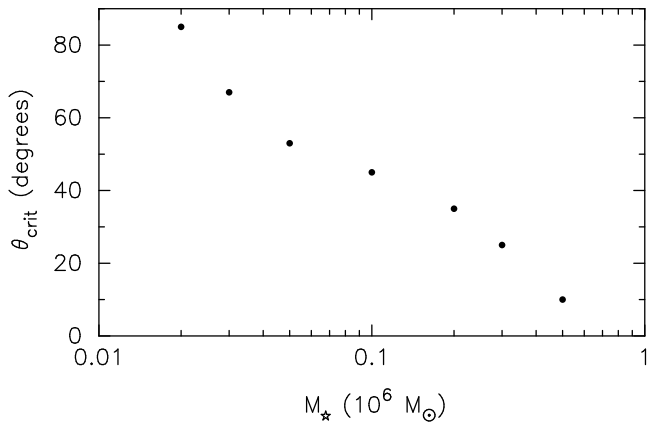}
\caption{\label{Figure:theta0} Critical value of the initial angle between $\boldsymbol{S}$
and $\boldsymbol{L}_\mathrm{tot}$ that separates the two evolutionary modes:
damped precession ($\theta_0<\theta_\mathrm{crit}$) and 
continued precession ($\theta_0>\theta_\mathrm{crit}$).
The other parameters of the initial models are given in the text.
}  
\end{figure}

Evolution of the second sort, or ``damped precession,'' which leads to almost complete alignment of \sbh\ spin with $\boldsymbol{L}_\mathrm{tot}$, is not excluded by the conservation laws 
(\ref{Equation:JConstraint}), (\ref{Equation:OtherConstraints}), and in principle
could occur for any initial conditions.
In practice, we found that it occurs only 
when $\theta_0$ is sufficiently small.
The critical angle, $\theta_\mathrm{crit}$, 
separating the two evolutionary modes was found to depend on the
other model parameters.
Figure~\ref{Figure:theta0} shows the dependence of $\theta_\mathrm{crit}$
on the mass of the stellar cluster, when the other initial parameters are the
same as in Figures~\ref{Figure:sphere40} and \ref{Figure:sphere70}.

A large number of such integrations revealed that the two modes of evolution
illustrated in Figures~\ref{Figure:sphere40} and~\ref{Figure:sphere70} are generic.
Roughly speaking, the system may end up in one of two distinct states:
\begin{itemize}
\item Aligned $\boldsymbol{S}, \boldsymbol{L}_\mathrm{tot}$ and $\boldsymbol{J}$;
\item Uniform precession of both $\boldsymbol{S}$ and $\boldsymbol{L}_\mathrm{tot}$ 
about a fixed axis, essentially the axis of total angular momentum 
$\boldsymbol{J}=\boldsymbol{S}+\boldsymbol{L}_\mathrm{tot}$.
\end{itemize}
In the latter case, typically the angle $\theta$ between $\boldsymbol{S}$ and 
$\boldsymbol{L}_\mathrm{tot}$ decreases from its initial value $\theta_0$, 
but settles at some non-zero average value after a couple of precessional periods. 
As noted above, the overall precession frequency  may be estimated 
as the Lense-Thirring time for stars at the radius such that the total angular momentum 
of stars within this radius is  equal to $|\boldsymbol{S}|$. 
This frequency depends only weakly on the angle $\theta_0$ 
provided that $\theta_0\ne 0$.

If the total angular momentum of the stars is less than $\boldsymbol{S}$, 
 essentially all the orbital $\boldsymbol{L}_j$ end up aligned or counter-aligned with 
the \sbh\ spin, depending on whether 
$\theta_0$ is greater or less than $\pi/2$.
This situation is unlikely to be relevant for  galactic nuclei, 
since there will always be enough stars sufficiently far from SBH that their total angular momentum exceeds $\boldsymbol{S}$, although the precessional times associated
with distant stars may be long. 
In addition, if the stellar orbits are initially concentrated in a small interval of 
radii on nearly-circular orbits, that is, have very little scatter in their individual 
precession frequencies $\boldsymbol{\omega}_j$, steady precession without 
alignment may persist even for $\boldsymbol{L}_\mathrm{tot} \ll \boldsymbol{S}$.

\section{\label{Section:Encounters}Influence of gravitational encounters on \sbh\ spin}

Times associated with Lense-Thirring precession about a \sbh\ are long, 
and over such long time scales, stellar orbits can evolve in response to other influences.
Here we consider how (Newtonian) gravitational interactions between stars would alter the
evolution of \sbh\ spins.
These interactions are expected to be most important in the dense nuclei
of low-luminosity spheroids; we derive more exact criteria below.
We continue to assume that $\boldsymbol S$ evolves according to Eqs.~(\ref{Equation:SpinEvol}),
but we now allow for the possibility of other terms in the 
the evolution equations for the $\boldsymbol{L}_j$.

\subsection{\label{Section:EncountersTimeScales}Encounter time scales}

To a first approximation, the force from the stars can be modelled
by approximating their distribution as spherically symmetric and stationary.
The addition of a spherical component to the otherwise Keplerian potential
of the \sbh\ results in an advance of orbital periapsis of each star 
(``apsidal precession'') at an orbit-averaged rate given by \cite{DEGN}
\beq
\label{Equation:nuM}
\nu_\mathrm{M} = -\nu_r G_\mathrm{M}(e,\gamma)\left(1-e^2\right)^{1/2} 
\left[\frac{M_\star(a)}{\mh}\right]
\eeq
with associated time scale
\begin{eqnarray}\label{Equation:DefinetM}
t_\mathrm{M} \equiv \left|\frac{\pi}{\nu_\mathrm{M}}\right| \approx
\frac{P}{2}\left(1-e^2\right)^{-1/2}\left[\frac{\mh}{M_\star(a)}\right].
\end{eqnarray}
Here, $\nu_r\equiv 2\pi/P$ is the Keplerian (radial) frequency, 
$M_\star(a)$ is the mass in stars within radius
$r=a$, and $G_\mathrm{M}\approx 1$ is a weak function of $\gamma$ and $e$
\cite{DEGN}.
Adopting our parametrization for low-mass galaxies, this becomes
\begin{eqnarray}\label{Equation:DefinetMLow}
t_\mathrm{M} &\approx& \frac{1.5\times 10^5 }{\left(1-e^2\right)^{1/2}} 
\left(\frac{\mh}{M_{0.1}}\right)
\left(\frac{a}{0.1 \mathrm{pc}}\right)^{\gamma-3/2} \mathrm{yr} 
\end{eqnarray}
while for high-mass galaxies,
\begin{eqnarray}\label{Equation:DefinetMHigh}
t_\mathrm{M} &\approx& \frac{2.3\times 10^4}{\left(1-e^2\right)^{1/2}} 
\left(\frac{\rm}{1\,\mathrm{pc}}\right)^{3/2}
\left(\frac{a}{\rm}\right)^{\gamma-3/2} \mathrm{yr} .
\end{eqnarray}
This ``mass precession'' leaves the orbital plane, and hence $\boldsymbol{L}_j$, 
unchanged and so does not directly affect
the evolution of $\boldsymbol{S}$ as given by Eq.~(\ref{Equation:SpinEvol}).
The same is true for the in-plane precession  due to the Schwarzschild
and Kerr parts of the \sbh\ metric; in the orbit-averaged, post-Newtonian approximation, 
the time scale asociated with the former precession, which always dominates the
the Kerr contribution, is
\begin{eqnarray}\label{Equation:nuGR}
t_\mathrm{S} &\equiv& \left|\frac{\pi}{\nu_\mathrm{S}}\right| = 
\frac{\pi}{3} \frac{(1-e^2)a^{5/2}c^2}{\left(G\mh\right)^{3/2}}\nonumber \\
&\approx& 1.0\times 10^9 \left(1-e^2\right)
\left(\frac{a}{0.1 \mathrm{pc}}\right)^{5/2}
\left(\frac{\mh}{10^6\msun}\right)^{-3/2} \mathrm{yr} . \nonumber \\
\end{eqnarray}

This ``Schwarzschild precession'' is more rapid than mass precession
when 
\beq
\left(1-e^2\right)^{3/2}\left(\frac{a}{a_\mathrm{S}}\right) < 1
\eeq
where
\beq\label{Equation:DefineaS}
a_\mathrm{S} M_\star(r<a_\mathrm{S}) = 3\mh \rg .
\eeq
For low-mass galaxies this is
\beq\label{Equation:DefineaSLow}
\left(\frac{a_\mathrm{S}}{0.1 \mathrm{pc}}\right)^{4-\gamma} \approx
3\frac{\mh}{M_{0.1}}\frac{\rg}{0.1\,\mathrm{pc}}
\eeq
and for high-mass galaxies,
\beq\label{Equation:DefineaSHigh}
\left(\frac{a_\mathrm{S}}{\rm}\right)^{4-\gamma} \approx
\frac32 \frac{\rg}{\rm}.
\eeq
For example, setting $\gamma=2$ in the first relation gives
\begin{equation}
a_\mathrm{S} \approx 1.2
\left(\frac{\mh}{10^6\msun}\right)
 \left(\frac{M_{0.1}}{10^4\msun}\right)^{-1/2} \, \mathrm{mpc} .
\end{equation}
While not directly affecting the $\boldsymbol{L}_j$, these two sources of
precession are important in setting the time scale for random fluctuations in the
orbital eccentricities, as discussed in more detail below.

Newtonian perturbations can also mimic frame-dragging by changing the orientation of orbital planes.
If such changes occur on a time scale that is short compared with the Lense-Thirring
precessional time, the evolution of orbital orientations will be determined essentially by the Newtonian perturbations \cite{MAMW2010}.
We expect this to be the case for stars that are sufficiently far from the \sbh, 
since frame-dragging time scales
increase rapidly with distance (Eq.~\ref{Equation:OmegaKb}).

Here we focus on a generic source of non-spherically-symmetric perturbations:
resonant relaxation (RR), the changes in $\boldsymbol{L}$ that result from
the finite-$N$ asymmetries in an otherwise spherical cluster around
a \sbh\ \cite{RauchTremaine1996}.
(Other possible sources of non-sphericity, ignored here, include a large-scale
distortion of the nuclear potential or ``bar'' \cite{MerrittVasiliev2010},
or a distant massive perturber \cite{Perets2007}.)
In what follows, we call the evolution of orbital planes due to these mutual torques
``2d resonant relaxation,'' or 2dRR 
\footnote{Another common name is ``vector resonant relaxation.'' We are following the nomenclature of reference \cite{DEGN}.}.

Under 2dRR, orbital orientations change in a characteristic time \cite{DEGN}
\begin{eqnarray}\label{Equation:T2dRR}
T_\mathrm{2dRR} &\approx& \frac{P}{2\pi} \frac{\mh}{m_\star}\frac{1}{\sqrt{N}}\\
&\approx& 4.7\times 10^4 \left(\frac{a}{\mathrm{mpc}}\right)^{3/2} 
\left(\frac{\mh}{10^6\msun}\right)^{-1/2} \nonumber \\ &\times& 
\left(\frac{\mh}{10^6 m_\star}\right)
\left(\frac{N}{10^2}\right)^{-1/2} \mathrm{yr}
\nonumber
\end{eqnarray}
where $P=P(a)=2\pi/\nu_r$ is the radial (Kepler) period and $N=N(a)$ is the
number of stars at $r\lap a$.
In a time $\sim T_\mathrm{2dRR}$, orbital planes will have essentially randomized
due to the mutual torques \footnote{This is also the time scale associated with
changes in eccentricity in the ``coherent resonant relaxation'' regime (Appendix).}. 

The condition that frame dragging causes orbital planes to precess
more rapidly than they are changed by  the mutual torques is
\beq
t_\mathrm{K}\equiv \frac{\pi}{\omega_\mathrm{LT}} \lap T_\mathrm{2dRR}
\eeq
or equivalently \footnote{This is essentially Eq.~(16) of Ref. \cite{MAMW2010},
and $a_\mathrm{K}$ is essentially $r_\mathrm{crit}$ from that paper.}
\beq\label{Equation:KerrCondition}
\left(1-e^2\right)^3\left(\frac{a}{\rg}\right)^3 \lap \frac{16\chi^2}{N(a)}
\left(\frac{\mh}{m_\star}\right)^2.
\eeq
Orbits satisfying this condition will be said to be in the ``collisionless''
regime: to a first
approximation, their angular momenta evolve in accordance with
 Eq.~(\ref{Equation:Ljdot}), unaffected by perturbations from other stars.

The condition~(\ref{Equation:KerrCondition})
can be expressed in terms of a characteristic semimajor axis,
 $a_\mathrm{K}$, as
\begin{equation}\label{Equation:DefineaK}
\left(1-e^2\right)^3\left(\frac{a}{a_\mathrm{K}}\right)^{6-\gamma} \lap 1.
\end{equation}
We call $a_\mathrm{K}$ the ``rotational influence radius'' of the \sbh.
To solve for $a_\mathrm{K}$, we write $N(a)$ for each of the
two types of nuclear model defined in \S\ref{Section:Models}
as
\begin{subequations}
\begin{eqnarray}
N(a) &\approx& \frac{M_{0.1}}{m_\star}\left(\frac{a}{0.1\,\mathrm{pc}}\right)^{3-\gamma},
\ \ \mh\lap 10^{7.5}\mh \label{Equation:NofaLowMass}\\
N(a) &\approx& 2\frac{\mh}{m_\star} \left(\frac{a}{\rm}\right)^{3-\gamma}, \ \ 
\mh\gap 10^{7.5}\msun. \label{Equation:NofaHighMass}
\end{eqnarray}
\end{subequations}
These approximate expressions are adequate given the approximate nature
of Eq. (\ref{Equation:KerrCondition}).
In the case of low-mass galaxies, Eqs. (\ref{Equation:KerrCondition})
- (\ref{Equation:NofaLowMass}) yield
\begin{eqnarray}\label{Equation:DefineaKLowb}
&&\left(\frac{a_\mathrm{K}}{0.1\,\mathrm{pc}}\right)^{6-\gamma} \approx
1.8\times 10^{-11}\,\chi^2 \\
&\times&\left(\frac{\mh}{10^6\msun}\right)^5 
\left(\frac{M_{0.1}}{10^5\msun}\right)^{-1} 
\left(\frac{m_\star}{1 \msun}\right)^{-1} \nonumber
\end{eqnarray}
while for high-mass galaxies, Eqs. (\ref{Equation:rmvsmh}), 
(\ref{Equation:KerrCondition}) - (\ref{Equation:DefineaK})
and (\ref{Equation:NofaHighMass}) give
\begin{eqnarray}\label{Equation:DefineaKHighb}
\left(\frac{a_\mathrm{K}}{35\,\mathrm{pc}}\right)^{6-\gamma} &\approx&
2.1\times 10^{-12}\,\chi^2  \\
&\times& \left(\frac{\mh}{10^8\msun}\right)^{4+\alpha(3-\gamma)} 
\left(\frac{m_\star}{1 \msun}\right)^{-1} \nonumber  
\end{eqnarray}
with $\alpha \approx 0.56$.
Figure~\ref{Figure:Radii} plots $a_\mathrm{K}$ as a function of nuclear parameters.
In low-mass galaxies, $a_\mathrm{K}\ll a_\mathrm{L}$; 
as $\mh$ increases, $a_\mathrm{K}$ can approach $a_\mathrm{L}$.
In the latter case, we expect the net angular momentum associated with
stars inside the rotational influence sphere to be comparable with $S$.
\begin{figure}[h!]
\includegraphics[width=4.3cm,angle=-90.]{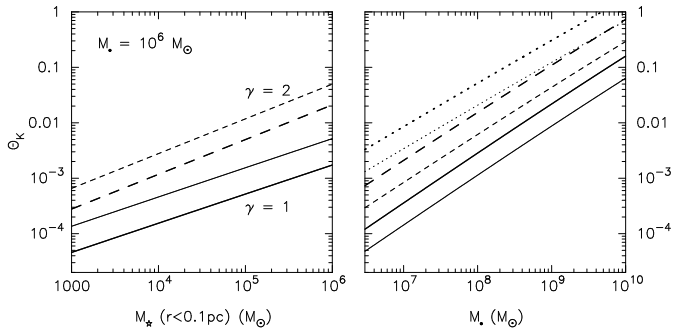}
\caption{\label{Figure:Theta} $\Theta_\mathrm{K}$ is the 
angular momentum associated with stars in the collisionless
regime, expressed as a fraction of the \sbh\ spin $S=\chi G\mh^2/c$.
The left panel assumes $\mh=10^6\msun$, $\chi=1$ and $f=1/2$
(maximal rotation of \sbh\ and stellar cluster); 
the stellar density is parametrized
in terms of $M_{0.1}$, the mass within $0.1$ pc, and $\gamma$, the power-law
index; solid lines are for $\gamma=1$ and dashed lines for $\gamma=2$.
Two values are assumed for the stellar mass: $m_\star=1\msun$ (thin lines) and $m_\star=10\msun$ (thick lines).
The right panel, for massive galaxies, assumes the relation
 (\ref{Equation:rmvsmh}) between $\mh$ and the influence radius $\rm$;
solid, dashed and dotted lines are for $\gamma=5/8$, $1$ and $3/2$ respectively.
Thick lines assume $\chi=1$, $f=0.5$ (rapidly rotating cluster)
and  thin lines assume $\chi=1$, $f=0.2$ (slowly rotating cluster).
All curves in the right panel assume $m_\star = 1\msun$.
}
\end{figure}

Define 
\beq
\Theta_\mathrm{K} \equiv \frac{L_\mathrm{K}}{S}
\eeq
where $L_\mathrm{K}$ is the angular momentum associated with stars
that satisfy (\ref{Equation:DefineaK}).
We compute $L_\mathrm{K}$ from Eq. (\ref{Equation:LtotIntegralb}) after
modifying the integration limits to respect the condition (\ref{Equation:DefineaK}).
The result is
\beq\label{Equation:ThetaKBright}
\Theta_\mathrm{K} = H(\gamma)  \left(\frac{f}{\chi}\right)
\sqrt{\frac{a_\mathrm{K}}{\rg}} \left(\frac{a_\mathrm{K}}{\rm}\right)^{3-\gamma}
\eeq
for bright galaxies, and
\beq\label{Equation:ThetaKFaint}
\Theta_\mathrm{K} = \frac12 H(\gamma)  \left(\frac{f}{\chi}\right)
\sqrt{\frac{a_\mathrm{K}}{\rg}} \left(\frac{M_{0.1}}{\mh}\right)
\left(\frac{a_\mathrm{K}}{0.1\,\mathrm{pc}}\right)^{3-\gamma}
\eeq
for faint galaxies, where
\begin{eqnarray}
&&H(\gamma) \equiv \\
&& \frac83 \frac{\sqrt{\pi}}{2^\gamma} \frac{(3-\gamma)}{(\gamma-1)}
\frac{\Gamma(\gamma+1)}{\Gamma(\gamma-1/2)}
\left[\frac{6-\gamma}{7-2\gamma} - \left(\frac{a_\mathrm{max}}{a_\mathrm{K}}\right)^{(1-\gamma)/2}\right]  \nonumber
\end{eqnarray}
for $\gamma\ne 1$, and
\begin{equation}
H(1) = \frac{8}{15}  +  \frac43\log\left(\frac{a_\mathrm{max}}{a_\mathrm{K}}\right).
 \nonumber
\end{equation}
(An upper cutoff to $a$ is only required when $\gamma \le 1$ due to a weak
divergence of the integral;
a natural choice is $a_\mathrm{max}=\rm$ since the
expressions for $N(a,e)$, etc. are only valid at $r<\rm$.)

Figure~\ref{Figure:Theta} plots $\Theta_\mathrm{K}$ as a function
of nuclear parameters.
As expected, for high-mass galaxies, and for $f\approx 1/2$, 
$\Theta_\mathrm{K}$ is of order unity, scaling as 
$f\chi^{-\gamma/(6-\gamma)}$ for fixed $\mh$.
As $\mh$ is decreased,  $\Theta_\mathrm{K}$ falls as well, 
although it can still be appreciable, $0.01\lap \Theta_\mathrm{K}\lap 0.1$,
in low-mass galaxies with dense nuclei, $\gamma\approx 2$. 

Figure \ref{Figure:Theta} suggests that in galaxies with large $\mh$, 
the joint evolution of $\boldsymbol{S}$ and $\boldsymbol{L}_j$
 will be similar to the evolution described in the previous section,
in the sense that mutual stellar interactions can be neglected.
As $\mh$ is decreased, the angular momentum associated with stars in the
collisionless regime drops compared with $S$.
In nuclei with sufficiently small $\mh$, 
most of the torque acting on the \sbh\ is likely to originate
in stars whose orbits respond to each other on a  shorter
time scale than the local Lense-Thirring time.
As a result, the angular momentum vectors of these stars will be unable to align 
around $\boldsymbol{S}$ as in the collisionless case.
We argue in the next section that the result can be substantially higher
rates of sustained \sbh\ precession.

\subsection{\label{Section:EncountersOU}Stochastic model for the evolution of $\boldsymbol{\omega}_S$}

In principle, the combined effects of gravitational self-interactions
and spin-orbit torques
could be directly simulated using an $N$-body algorithm
\cite{MAMW2010}.
However the ratio between Kerr precessional times and orbital periods is
so great
that such direct simulation would be  expensive for any reasonable $N$.

An alternative approach would be to incorporate the effects of star-star
interactions by modeling the evolution of each of the
$\boldsymbol{L}_j$ as a random walk \cite{Eilon2009,MAMW2011,Madigan2011}.
However, interactions between stars must conserve $\boldsymbol{L}_\mathrm{tot}$,
as well as being constrained in less obvious ways by the fact that the torques
are mutual.
Approximating the evolution of each star's angular momentum as an independent 
stochastic process, independent of  the changes in the other $\boldsymbol{L}_j$, 
would fail to capture these essential constraints.

Since the effects of the $\boldsymbol{L}_j$ on ${\boldsymbol S}$ appear only through
$\boldsymbol{\omega}_S$, and since the time scales for changes in the $\boldsymbol{L}_j$
due to self-interactions are typically short compared with spin-orbit time scales,
it is reasonable to separate the problem into two parts:
asking first how $\boldsymbol{\omega}_S$ varies as the stars interact with one another, ignoring the effects of spin-orbit torques; then using this knowledge to predict
how $\boldsymbol{S}$ would evolve in response to the fluctuating $\boldsymbol{\omega}_S$.
We first explore this model, then present a more careful justification below.

\begin{figure}
\includegraphics[width=7.cm,angle=-90.]{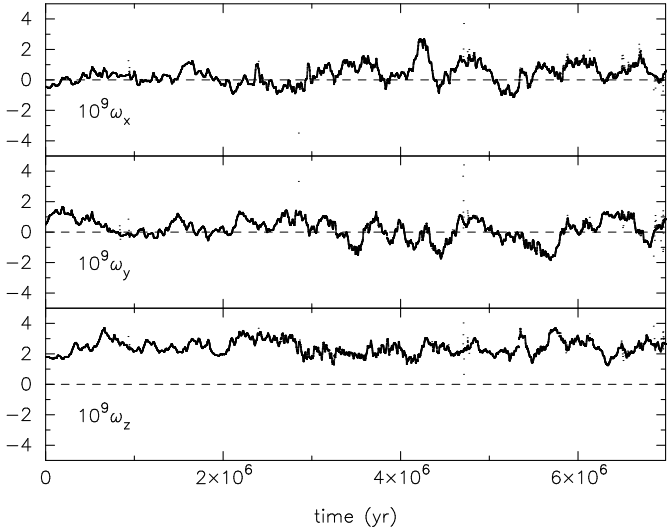}
\caption{\label{Figure:omega42} 
Evolution of $\boldsymbol{\omega}_S$ in the $N$-body integration described
in the text.
Units of $\omega$ are inverse years.}
\end{figure}

Figure~\ref{Figure:omega42}, based on a direct integration of the
$N$-body equations of motion for $100$ point masses (stars) orbiting 
about a massive particle (\sbh), illustrates how $\boldsymbol{\omega}_S$ evolves
due to star-star interactions.
The integrator \cite{MikkolaMerritt2006,MikkolaMerritt2008} 
included 1PN terms in \sbh-star
interactions; spin-orbit terms were omitted.
Initial conditions were generated according to Eq.~(\ref{Equation:Nofae}),
with $\mh=10^6\msun$, $m_\star=10\msun$, $a_\mathrm{max}=10$ mpc, $r_{p,\mathrm{min}}=1$ mpc, $\gamma=1$ and $f=1/2$.
Rotation of the cluster was initially about the $z$-axis.

\begin{figure}
\includegraphics[width=7.cm,angle=-90.]{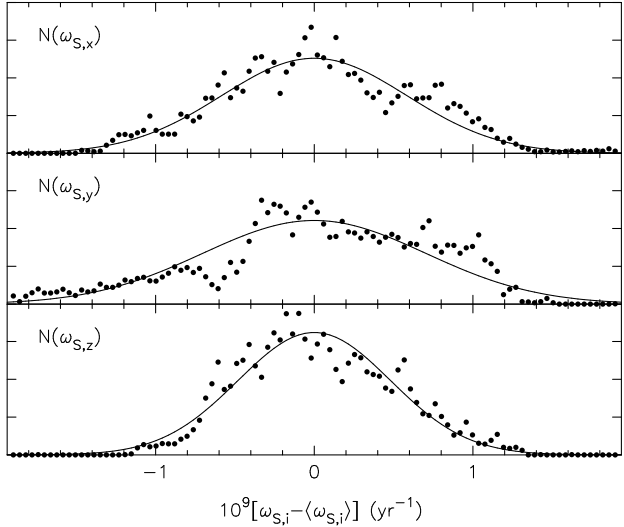}
\caption{\label{Figure:Fluc42} 
Distribution of the $\boldsymbol{\omega}_S$ values measured 
 in the $N$-body simulation of Figure~\ref{Figure:omega42}.
Solid lines are Gaussian fits with $\sigma_{x,y,z}=\{5.8,7.0,4.8\}\times 10^{-10}$
yr$^{-1}$.}
\end{figure}

Total angular momentum, $\boldsymbol{L}_\mathrm{tot}$ 
(or rather, its 1PN analog \cite{DD-85}) 
is  conserved in these $N$-body integrations.
The spin precessional vector
 is not conserved;
but since $\boldsymbol{\omega}_S$ is a weighted sum of the $\boldsymbol{L}_j$, 
and since $\sum\boldsymbol{L}_j$ is conserved,
exchange of angular momentum between stars tends on average to leave $\boldsymbol{\omega}_S$
unchanged.
However, Figure~\ref{Figure:omega42} shows that each component of $\boldsymbol{\omega}_S$ fluctuates about its mean value in an apparently 
random fashion.
The amplitude of these fluctuations is approximately constant over time, 
giving each time series the appearance of a stationary stochastic process
\cite{VanKampen1992}.

Assuming stationarity, it is reasonable to calculate the distribution function of the
fluctuations at any given time by binning together the events from all times.
The results are shown in Figure~\ref{Figure:Fluc42} where the distributions
have been fit to Gaussian functions.

\begin{figure}
\includegraphics[width=7.cm,angle=-90.]{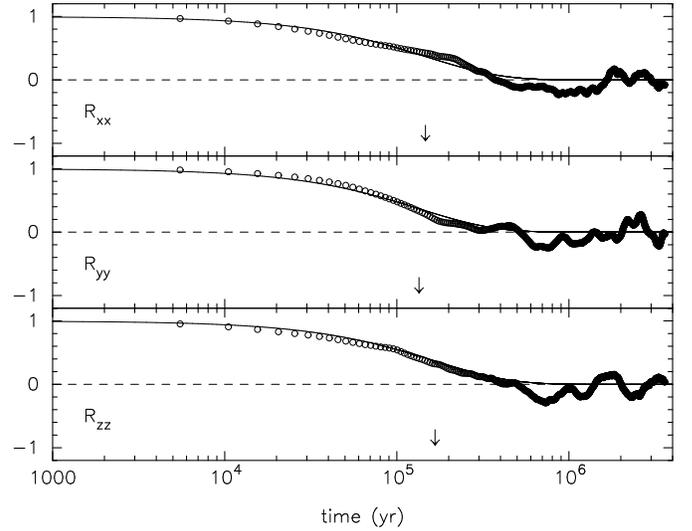}
\caption{\label{Figure:acf42} 
Circles are auto-correlation functions of the time series plotted in Figure~\ref{Figure:omega42};
horizontal axis is the lag time.
Arrows mark the computed correlation times.
Solid lines are  Eq. (\ref{Equation:RiiOU}).}
\end{figure}

The time scale associated with stochastic  fluctuations in  $\boldsymbol{\omega}_S$
in the $N$-body integrations can be found by computing the
autocorrelation functions (ACF), defined as
\beq
R_{ii}(t)  = \frac{\int_0^T \left[\omega_{S,i}(t^\prime)-\overline{\omega}_{S,i}\right]\,\left[\omega_{S,i}(t^\prime+t)-\overline{\omega}_{S,i}\right]\, dt^\prime}
{\int_0^T [\omega_{S,i}(t^\prime) - \overline{\omega}_{S,i}]^2 dt^\prime}.
\eeq
Here, $\omega_{S,i}$ is the $i$th component of $\boldsymbol{\omega}_S$,
$\overline{\omega}_{S,i}$ is its time-averaged value,
and $0\le t\le T$ is the elapsed time in the $N$-body integration.
Figure~\ref{Figure:acf42} shows that 
the measured ACF's are reasonably well fit by exponential functions:
\beq\label{Equation:RiiOU}
R_{ii}(t) \approx  \exp(-t/\tau_i), \ \ \ \ i = \{1,2,3\},
\eeq
with $\tau_i\approx 1.5\times 10^5$ yr.

One expects the autocorrelation time for $\boldsymbol{\omega}_S(t)$
to be similar to the characteristic time associated with changes in the $\boldsymbol{L}_j$.
One such time is the 2d resonant relaxation time defined in Eq.~(\ref{Equation:T2dRR}).
There will also be variations in $\boldsymbol{\omega}_\mathrm{S}$ due to changes
in orbital eccentricities.
In the so-called ``coherent RR'' regime, defined as 
$\Delta t \lap \{t_\mathrm{M}, t_\mathrm{S}\}$,
changes in $e$ occur in a characteristic time $\sim T_\mathrm{2dRR}$;
while in the ``incoherent RR'' regime, i.e.
$\Delta t \gap \{t_\mathrm{M}, t_\mathrm{S}\}$,
the associated time  is longer than $T_\mathrm{2dRR}$ (Appendix).
Hence $T_\mathrm{2dRR}$ is the shortest of the variability time scales and
we can safely associate $\tau$ with it.
For the $N$-body models, Eq.~(\ref{Equation:T2dRR}) states
\beq
T_\mathrm{2dRR}\approx 5\times 10^4 \left(\frac{a}{1\,\mathrm{mpc}}\right) \mathrm{yr},
\eeq
quite consistent with the autocorrelation times measured in the $N$-body simulation
given that $1\,\mathrm{mpc}\lap a \lap 10\,\mathrm{mpc}$.
The longer time scale associated with randomization of orbital eccentricities
is given by Eq.~(\ref{Equation:TRRM}):
\begin{equation}\label{Equation:TRRMNbody}
T_\mathrm{RR,M}(a) 
\approx 3\times 10^6 \left(\frac{a}{1\, \mathrm{mpc}}\right)^{3/2} \mathrm{yr}
\end{equation}
for $a\gap a_\mathrm{S}\approx 0.00$ mpc.

The variance in the components of $\boldsymbol{\omega}_S$,
\begin{equation}
\sigma_i^2 \equiv \overline{[\omega_{S,i}(t) - \overline{\omega}_{S,i}]^2},
\end{equation}
can also be estimated given the known properties of the initial model.
Begin by rewriting $\boldsymbol{\omega}_S$, 
Eq.~(\ref{Equation:SpinEvol}b), as
\begin{subequations}
\begin{eqnarray}
\omega_{S,i}(t) &=& \sum_j C_j \lambda_{j,i} , \\
C_j &=& \frac{2G^{3/2}\mh^{1/2}m_j}{c^2 a_{j}^{5/2}\left(1-e_j^2\right)} ,\\
\lambda_{j,i}(t) &=& \left[\boldsymbol{u}_{L,j}(t)\cdot \boldsymbol{u}_i \right]
\end{eqnarray}
\end{subequations}
with $\boldsymbol{u}_{L,j}(t)$ a unit vector in the direction of $\boldsymbol{L}_j$
and $\boldsymbol{u}_i$ a unit vector in the direction of the $i$th coordinate axis.
In general, each of the variables $\{e_j, \lambda_{j,i}\}$ will change stochastically
due to star-star interactions, and all of these changes will contribute to the variance
of $\boldsymbol{\omega}_S$.
A lower limit on that variance follows from assuming that resonant relaxation
causes changes only in the orbital planes and that 
the $e_j$ are approximately constant.
Then
\begin{subequations}
\begin{eqnarray}
\sigma_i^2 \equiv 
\mathrm{var}\left[\omega_{S,i}\right] &\approx& \sum_j C_j^2 \times \mathrm{var}\left[\lambda_{j,i}\right] \\
&\approx& \sum_j C_j^2\\
&\approx& 
\sum_j \frac{4 G^3\mh m_j^2}{c^4  a_j^{5}\left(1-e_j^2\right)^{2}} 
\end{eqnarray}
\end{subequations}
assuming $\mathrm{var}[\lambda_j]\approx 1$.
For a cluster containing orbits with a single $(a,e)$, the right hand side
is $\sim \omega_{S,\mathrm{max}}^2/N$, where $\omega_{S,\mathrm{max}}$
is the magnitude of $\boldsymbol{\omega}_S$ in a maximally-rotating cluster
with $f=1/2$.
Then $\sigma\approx \omega_{S,\mathrm{max}}/\sqrt{N}$.

According to Eq.~(\ref{Equation:TRRMNbody}), orbital eccentricities should
also change substantially over the integration period, particularly for orbits
of small $a$.
The variance in $\boldsymbol{\omega}_S$ should therefore
contain a substantial contribution
from changes in the $e_j$, 
and in fact the formula just derived underpredicts
the variances observed in the $N$-body simulations (Figure~\ref{Figure:Fluc42})
by a factor of a few.
We estimate $\sigma$ allowing for changes in the $e_j$ as follows.
Rewrite Eq.~(\ref{Equation:SpinEvol}b) yet again as
\begin{subequations}
\begin{eqnarray}
\omega_{S,i}(t) &=& \sum_j A_j X_{j,i}(t), \\
A_j &=& \frac{2G^{3/2}\mh^{1/2}m_j}{c^2 a_{j}^{5/2}} , \\
X_{j,i} &=& \frac{\lambda_{j,i}}{1-e_j^2} 
\end{eqnarray}
\end{subequations}
where both $\boldsymbol{\lambda}_j$ and $e_j$ are allowed to be functions of time.
Assuming uncorrelated changes, the variance in $\omega_{S,i}$ is
\begin{subequations}\label{Equation:sigmaN}
\begin{eqnarray}\label{Equation:sigmaNa}
\sigma_i^2 &=& \sum_j A_j^2 \mathrm{var}\left[X_j\right], \\
\mathrm{var}\left[X_j\right] &=& \left(\overline{\lambda}_{j,i}\right)^2\mathrm{var}\left[(1-e_j^2)^{-1}\right] \nonumber \\
&+& \overline{(1-e_j^2)^{-1}}\, \mathrm{var}\left[\lambda_{j,i}\right]  \nonumber \\
&+& \mathrm{var}\left[(1-e_j^2)^{-1}\right] \mathrm{var}\left[\lambda_{j,i}\right].
\label{Equation:sigmaNb}
\end{eqnarray}
\end{subequations}
We estimate the quantities on the right hand side of Eq.~(\ref{Equation:sigmaNb}) by
assuming that resonant relaxation maintains a ``thermal'' distribution of 
eccentricities at every $a$ \cite{MAMW2011}, i.e. that
\beq
N(e) de \approx 2 e de
\eeq
for $0\le e \le e_\mathrm{max}(a)$, $e_\mathrm{max}\lap 1$. 
Then
\begin{eqnarray}
\overline{(1-e_j^2)^{-1}} &\approx& \ln\left[\left(1-e_\mathrm{max}^2\right)^{-1}\right], \\
\mathrm{var}\left[(1-e_j^2)^{-1}\right] &\approx& \left(1-e_\mathrm{max}^2\right)^{-1} - 1 \nonumber \\
&-& \left[\ln\left(1-e_\mathrm{max}^2\right)\right]^2 .
\end{eqnarray}
We identify $e_\mathrm{max}(a)$ with $1-r_{p,\mathrm{min}}/a$.
We likewise assume that orbital planes are randomized, so that
$\mathrm{var}[\lambda_{j,i}]=1/3$, and $\overline{\lambda}_{j,i}=f$
in the case that $u_i\parallel \boldsymbol{L}_\mathrm{tot}$ and zero otherwise.
Finally, since $e_\mathrm{max}\approx 1$, it is reasonable to ignore 
the logarithmic terms, yielding
\beq\label{Equation:sigmaest}
\sigma_i^2 \approx \frac23\frac{G^3 \mh}{c^4}
\sum_j \frac{m_j^2}{a_j^5}
\left(\frac{r_{p,\mathrm{min}}}{a_j}\right)
\left(\frac{a_j}{r_{p,\mathrm{min}}}-1\right)^2
\eeq
for  $i = 1, 2, 3$.

Applied to the $N$-body models in Figure~\ref{Figure:Fluc42}, Eq.~(\ref{Equation:sigmaest})
yields $\sigma_i \approx 6\times 10^{-10}$ yr$^{-1}$, in good agreement with the values
obtained via the Gaussian fits to the $N$-body data.

As long as the characteristic time for changes in eccentricity is shorter than the
other times of interest, Eq. (\ref{Equation:sigmaest}) is the appropriate expression to use
for $\sigma$. This will turn out always to be the case in the examples presented below.

A theorem \cite{VanKampen1992} states that a stationary random process
with a Gaussian probability function and an exponentially decaying 
autocorrelation function is necessarily an Ornstein-Uhlenbeck \cite{Uhlenbeck1930} 
process.
The latter is defined as having a transition probability between two states,
$y_1$ and $y_2$ (given here by two values of $\boldsymbol{\omega}_S$), at times $t_1$ and
$t_2$ that obeys
\beq
T(y_2|y_1) = \frac{1}{\sqrt{2\pi\left(1-e^{-2\Delta}\right)}}\exp\left[-\frac{(y_2-y_1e^{-\Delta})^2}{2(1-e^{-2\Delta})}\right]
\eeq
where $\Delta=(t_2-t_1)/\tau$ and $\tau$ is defined as in Eq.~(\ref{Equation:RiiOU}).
An OU process $X(t)$ with mean value $\overline{X}$ 
can also be defined as the solution of the Langevin equation,
\beq
\dot X = - \gamma \left[X(t) - \overline{X}\right] + {\cal N}(t)
\eeq
if $\gamma=\tau^{-1}$ and if ${\cal N}(t)$ is a Gaussian
random variable having the properties
\begin{subequations}
\begin{eqnarray}
\overline{{\cal N}(t)} &=& 0, \\
\overline{{\cal N}(t){\cal N}(t')} &=& \Gamma\delta(t-t')
\end{eqnarray}
\end{subequations}
with $\Gamma = 2\sigma^2/\tau$ \cite{VanKampen1992}.
This comparison suggests that $\boldsymbol{\omega}_S$ experiences
a ``frictional force,'' of amplitude $-(\boldsymbol{\omega}_S-\overline{\boldsymbol{\omega}}_S)/\tau$, that tends
to bring that vector back to its original value in spite of the fluctuations.
This ``force'' is presumably related to the physical constraint $\boldsymbol{L}_\mathrm{tot}=\mathrm{const.}$, although we do not 
explore the nature of that connection here.

\begin{figure}
\includegraphics[width=6.cm,angle=-90.]{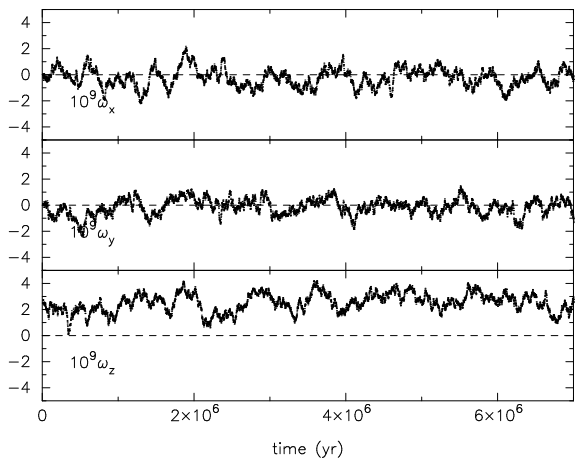}
\caption{\label{Figure:ou42} 
Ornstein-Uhlenbeck realization of $\boldsymbol{\omega}_S(t)$ using 
($\tau,\sigma$) derived from the $N$-body simulations in Figure~\ref{Figure:omega42}.}
\end{figure}

A stochastic realization of an OU process $X(t)$ can be generated via  \cite{Gillespie1996}
\begin{eqnarray}\label{Equation:OU2}
X\left(t+\Delta t\right) -\overline{X} &=& \left[X(t) - \overline{X}\right]e^{-\Delta t/\tau} \nonumber \\
&+& 
\left[\sigma^2\left(1-e^{-2\Delta t/\tau}\right)\right]^{1/2} n
\end{eqnarray}
where $n$ is a sample value of the unit normal random variable,
and, in our case, $X(t)$ is one of the components of $\boldsymbol{\omega}_S$.
Figure~\ref{Figure:ou42} shows an example generated 
from Eq.~(\ref{Equation:OU2})
using the values of $\{\tau_i,\sigma_i\}$ extracted from the 
$N$-body simulation data of Figure~\ref{Figure:omega42}.
For this example, $\overline{\omega}_x$ and $\overline{\omega}_y$
were zero (rotation of the cluster about the $z$-axis)
and $\overline{\omega}_z$ was set to its initial value.
 
We can use these results to rewrite Eqs.~(\ref{Equation:SpinEvol})
and~(\ref{Equation:Ljdot}) in an approximate way that incorporates the
effects of star-star interactions.
Orbits that satisfy the condition (\ref{Equation:KerrCondition}) at $t=0$ are
assumed to evolve, collisionlessly, in response to spin-orbit torques,
according to Eq.~(\ref{Equation:Ljdot}), and the contribution of these stars
to $\boldsymbol{\omega}_S$, which we call $\boldsymbol{\omega}_K$, is computed as in
Eq.~(\ref{Equation:SpinEvol}b).
In the case of orbits that do not satisfy (\ref{Equation:KerrCondition}), no
attempt is made to follow their detailed evolution.
Instead, these orbits are assumed to make a collective, stochastic contribution
to $\boldsymbol{\omega}_S$ which is modeled as an Ornstein-Uhlenbeck
 time series, $[\boldsymbol{\omega}_S]_\mathrm{OU}$, evaluated numerically
via Eq.~(\ref{Equation:OU2}).
The parameters ($\tau,\sigma$) that appear in that equation are estimated
as described above.
These two contributions to $\boldsymbol{\omega}_S(t)$ are then added,
and the evolution equation for $\boldsymbol{S}$  is written as
\begin{subequations}\label{Equation:SpinEvolOU}
\begin{eqnarray}\label{Equation:SpinEvolOUa}
\dot{\boldsymbol{S}} &=& \boldsymbol{\omega}_S\times \boldsymbol{S},\\
\boldsymbol{\omega}_S &=& \boldsymbol{\omega}_K + [\boldsymbol{\omega}_S]_\mathrm{OU}.\label{Equation:SpinEvolOUb}
\end{eqnarray}
\end{subequations}

Eq. (\ref{Equation:SpinEvolOU}) ignores the effects of frame-dragging on  stars
in the ``collisional'' region.
It therefore rules out the possibility that differential precession 
of stars in this region could distribute their
$\boldsymbol{L}_j$ vectors uniformly about $\boldsymbol{S}$,
causing their net torque on the \sbh\ to drop, as occurs in the collisionless regime
(Figure \ref{Figure:rpsdot}).
While we can not rigorously defend this approximation, 
we can state a more basic set of physical assumptions from which it follows.
Consider a star whose orbit evolves in response both to frame-dragging from
the \sbh\ and gravitational encounters from other stars.
Idealize the encounters as occurring at discrete times separated by 
$\sim T_\mathrm{RR}$.
Between encounters, the line of nodes precesses uniformly at the
Lense-Thirring rate, by an amount
\beq\label{Equation:DeltaOmegaFinite}
\Delta\Omega \approx \omega_\mathrm{LT} \times T_\mathrm{RR}.
\eeq
If the effect of an encounter is to randomly select a new $\Omega$ -- that is,
if memory of the previous $\Omega$ is completely erased after one relaxation time --
then the mean change in $\Omega$ after many encounters
will be just $\Delta\Omega$.
Finally, if $\omega_\mathrm{LT} T_\mathrm{RR}\ll 1$, then $\Delta\Omega\ll 2\pi$, implying a negligible amount of differential precession about $\boldsymbol{S}$
even after arbitrarily long times.

A similar argument \cite{Reif1965}
can be used to derive the drift velocity of an electron that
is subject to a fixed electric field (the \sbh\ torque) and to random collisions
(gravitational encounters); the finiteness of the drift velocity (nodal angle $\Omega$) 
follows from the assumption that collisions restore $\boldsymbol{v}$ to a thermal
distribution, i.e. that knowledge of the velocity accumulated prior to the collision
is lost.
As is well known, under some circumstances the charged particle retains memory
of the velocity it had before its collision leading to ``persistence-of-velocity''
corrections.
We expect our model for spin evolution to be similarly limited in its applicability,
although we postpone a more thorough understanding of such issues to a later
paper.

Some support for this physical picture is provided by Figure 2
of Merritt et al. \cite{MAMW2010}, which shows a set of short, numerical 
integrations of $N$-body systems subject to frame-dragging torques.
The lower left panel in that figure corresponds to the case
$T_\mathrm{RR}\approx \omega_\mathrm{LT}^{-1}$, and the lower right
panel to the case $T_\mathrm{RR}\ll \omega_\mathrm{LT}^{-1}$.
In the first case, stars exhibit a finite amount of nodal precession in spite
of the encounters,
as implied by Eq. (\ref{Equation:DeltaOmegaFinite}),
while in the latter case,  encounters appear to remove all traces of a net 
advance of $\Omega$.

Stars near the inner edge of the ``collisional'' region, $a\gap a_\mathrm{K}$, 
have $\omega_\mathrm{LT}\lap T_\mathrm{RR}^{-1}$,
and for these stars, $\Delta\Omega$ is not necessarily small, 
as in the lower-left panel of the figure just cited.
Given that the vaue of $a_\mathrm{K}$ is itself uncertain, the additional uncertainty
due to the evolution of orbits in this ``transition zone,'' 
$a_\mathrm{K}\lap r \ll a_\mathrm{L}$, seems acceptable.
We note that these uncertainties mimic uncertainties in twisted 
accretion disk models about the
location and radial extent of the ``warp'' that determines the torque
on the \sbh, as discussed in more detail in  \S\ref{Section:DiscussDisk}.
\begin{figure}[ht!]
\includegraphics[width=4.3cm,angle=-90.]{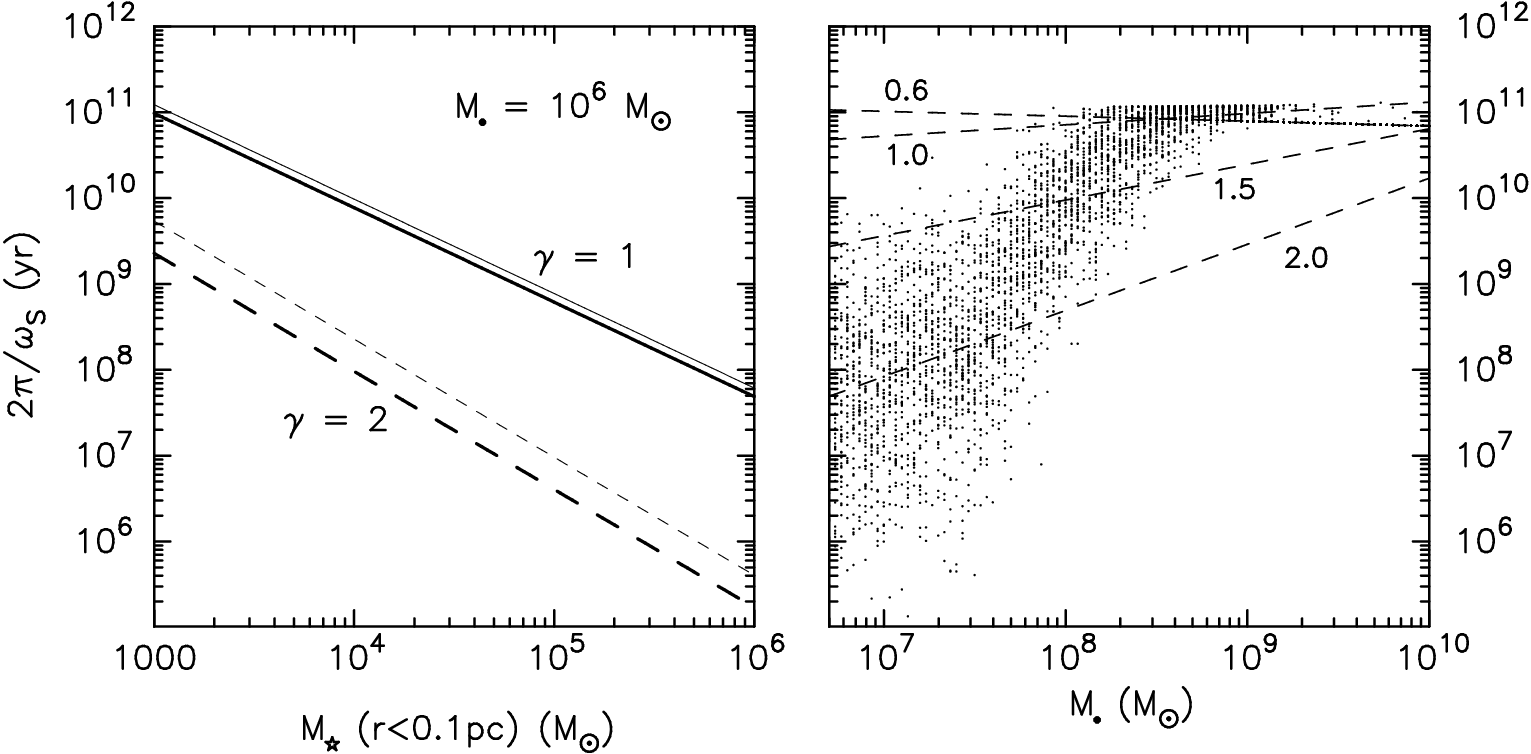}
\caption{\label{Figure:OmegaOU} 
\sbh\ precessional periods that would result from torquing by stars
that orbit outside the rotational influence sphere of the \sbh,
assuming $f=1/2$ and $\chi=1$.
In the left panel, $\mh=10^6\msun$ is assumed, and
the thin and thick lines correspond to 
$m_\star=1\msun$ and $m_\star=10\msun$ respectively.
The right panel assumes the relation (\ref{Equation:rmvsmh}) between
$\rm$ and $\mh$ and $m_\star = 1\msun$.
The points were computed from a Monte-Carlo model that approximates
the observed dependence of $\gamma$ on $\mh$; dashed lines are 
for constant $\gamma$, as labelled.
}
\end{figure}

In nuclei with $\Theta_\mathrm{K}\ll 1$ (Figure \ref{Figure:Theta}),
differential precession of the stars that contribute to $\omega_\mathrm{K}$ 
will cause their torque to die away before the direction
of $\boldsymbol{S}$ has changed appreciably (Figure \ref{Figure:rpsdot}).
Subsequent evolution of $\boldsymbol{S}$ will be 
determined by all the other stars.
In a nucleus described by the $N(a,e)$ of Eq. (\ref{Equation:Nofae}),
the contribution to $\omega_\mathrm{S}$ from those stars (ignoring
stochastic fluctuations) is given by
the integral (\ref{Equation:OmegaS}), after restricting the region of integration
to the complement of (\ref{Equation:DefineaK}).
The result is
\begin{subequations}\label{Equation:OmegaOU}
\begin{eqnarray}
\omega_\mathrm{S} &=& K(\gamma) f \frac{c}{\rg} \frac{M_0}{\mh}
\left(\frac{a_\mathrm{K}}{r_0}\right)^{3-\gamma}\left(\frac{a_\mathrm{K}}{\rg}\right)^{-5/2},\\
K(\gamma) &=& \frac{8}{3}\frac{\sqrt{\pi}}{2^\gamma}
\frac{(3-\gamma)(6-\gamma)}{(1-2\gamma)^2}
\frac{\Gamma(\gamma+1)}{\Gamma(\gamma-1/2)}
\end{eqnarray}
\end{subequations}
where $\{M_0,r_0\}=\{M_{0.1}, 0.1\,\mathrm{pc}\}$ in low-mass galaxies and
$\{M_0,r_0\}=\{2\mh,\rm\}$ in high-mass galaxies.

Figure ~\ref{Figure:OmegaOU} plots this contibution to the \sbh\ precessional period
 (that is, to the mean value of $2\pi/[\omega_\mathrm{S}]_\mathrm{OU}$) as a function
of nuclear parameters.
The results turn out to be strongly dependent on $\gamma$, the slope
of the nuclear density profile, so we consider that parameter in more detail.
Observationally, $\gamma$ exhibits a substantial scatter, but
there is a well-defined mean trend with galaxy luminosity, at least 
among the bright galaxies for which $\gamma$  is well-determined \cite{Gebhardt1996,ACSFornaxII}: $\gamma$ is smaller in the nuclei of brighter galaxies.
Using standard expressions for the mass-to-light ratio of old stellar
systems \cite{Worthey1994} and for the mean ratio of \sbh\ mass to
galaxy mass \cite{MarconiHunt2003}, we can write this mean relation as
\beq\label{Equation:GammaMean}
\langle\gamma\rangle \approx 2.0 - 1.1\log_{10}\left(\frac{\mh}{10^8\msun}\right),\ \ 
\mh \gap 10^{7.5}\mh.
\eeq
The right panel of Figure \ref{Figure:OmegaOU} shows a Monte-Carlo
distribution of points generated from this relation, assuming a dispersion
of $0.25$ in $\gamma$ at each $\mh$.
(Values of $\gamma\le 1/2$ were excluded for the reasons given above.)
While the scatter is large, there is also a steep trend in the sense
of smaller precessional periods at lower $\mh$.


\subsection{\label{Section:StochasticEvolution}Examples of stochastic evolution}

We first consider a dense nucleus in a low-mass galaxy, 
$M_\mathrm{gal}\approx 10^9\msun$:
we set  $\mh=10^6\msun$, $\gamma=2$ and $M_{0.1} = 2\times 10^5\msun$.
The characteristic radii relating to orbital coherence times
are given for this nuclear model 
by Eqs.~(\ref{Equation:DefineaSLow}) and (\ref{Equation:DefineaKLowb}):
\begin{subequations}  
\begin{eqnarray}
a_\mathrm{S} &\approx& 0.27\, \mathrm{mpc} \\
a_\mathrm{K} &\approx& 0.17 \chi^{1/2} \left(\frac{m_\star}{\msun}\right)^{-1/4} \mathrm{mpc}.
\end{eqnarray}
\end{subequations}
The total angular momentum associated with stars orbiting close enough to the 
\sbh\ that frame dragging dominates self-interactions, 
$a<a_\mathrm{K}(1-e^2)^{-3/4}$, is given 
by Eq. (\ref{Equation:ThetaKFaint}):
\beq
\Theta_\mathrm{K} \equiv \frac{L_\mathrm{K}}{S} \approx 0.018 \chi^{-1/4} 
\left(\frac{f}{0.5}\right)
\left(\frac{m_\star}{1\msun}\right)^{-3/8}.
\eeq
Evidently, an insignificant number stars are in the collisionless regime.
Almost all stars in the {\it collisional} regime will also have
$a\gap a_\mathrm{S}$; for these stars, 
the coherence time related to changes in eccentricity (Appendix) is given by Eq.~(\ref{Equation:DefinetMLow}):
\beq\label{Equation:tcohModel}
t_\mathrm{coh} \equiv t_\mathrm{M}(e=1/2) 
\approx 1.0\times 10^4 \left(\frac{a}{0.1\,\mathrm{pc}}\right)^{1/2} \mathrm{yr}.
\eeq
The incoherent, resonant relaxation time corresponding to this $t_\mathrm{coh}$,
Eq.~(\ref{Equation:TRRM}),  is then 
\beq
T_\mathrm{RR,M} \approx 3\times 10^6 \left(\frac{m_\star}{\msun}\right)^{-1}
\left(\frac{a}{1\,\mathrm{mpc}}\right)^{3/2} \mathrm{yr}
\eeq
which is the time associated with random changes in orbital eccentricities.
The 2d resonant relaxation time, in either the coherent or incoherent regimes,
is given by Eq.~(\ref{Equation:T2dRR}):
\beq\label{Equation:T2dRROU}
T_\mathrm{2dRR}\approx 1.1\times 10^4 
\left(\frac{m_\star}{M_\odot}\right)^{-1/2}
\left(\frac{a}{\mathrm{mpc}}\right) \mathrm{yr}
\eeq
which is the time scale associated with changes in orbital inclinations.
As expected, $T_\mathrm{2dRR}<T_\mathrm{RR,M}$.

The average, spin precessional period
of the \sbh\ due to torquing by stars in the collisional regime is 
given by Eq. (\ref{Equation:OmegaOU}):
\beq \label{Equation:tspinModel}
\frac{2\pi}{\omega_\mathrm{S}} \approx  3.7\times 10^6 
\left(\frac{f}{0.5}\right)^{-1} \chi^{3/4}
\left(\frac{m_\star}{1\,\msun}\right)^{-3/8} \mathrm{yr}.
\eeq 
Even assuming a low degree of net rotation of the cluster ($f\lap 0.1$), 
spin precessional periods are  predicted to be shorter than $\sim 10^8$ yr.

\begin{figure}[h!]
\includegraphics[width=7.5cm]{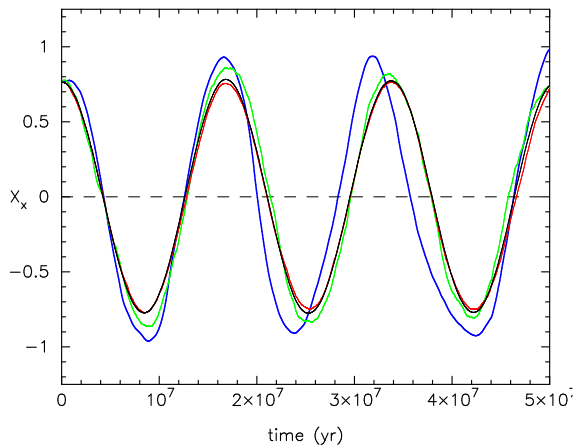}
\caption{\label{Figure:Tau} 
$x$-component of the spin vector $\boldsymbol{\chi}\equiv c\boldsymbol{S}/G\mh^2$,
derived as the solution to the stochastic differential equation (\ref{Equation:SpinEvolOU}),
with various values of $\tau$:
$\tau = 10^3$ yr (thinnest, black),
$10^4$ yr (red), $10^5$ yr (green), $10^6$ yr (thickest, blue).
Other parameters are $m_\star=1\msun$, $f=0.1$, $\theta_0=50^\circ$.}
\end{figure}


Since  $T_\mathrm{2dRR} \approx \tau \ll 2\pi/\omega_\mathrm{S}$,
we expect that the time evolution of $\boldsymbol{S}$ will depend only weakly 
on the value chosen for  $\tau$ in Eq. (\ref{Equation:OU2}),
as long as the inequality is maintained.
This expectation is confirmed in Figure \ref{Figure:Tau}, which shows
the evolution of the $x$-component of $\boldsymbol{S}$ in a set of
integrations with different $\tau$ and with $f=0.1$, $\theta_0=50^\circ$.
Only when $\tau$ is unphysically large, $\gap 10^6$ yr, and comparable
with the spin precessional period does $\boldsymbol{S}(t)$ show an appreciable
dependence on it.

\begin{table}[ht] 
\caption{Parameters for Figure~\ref{Figure:Series1}} 
\centering 
\begin{tabular}{c l c l c} 
\hline
\hline 
$m_\star/\msun$ & $N$ & $N_\mathrm{K}$ & $M_\mathrm{K}/\mh$ &
$a_\mathrm{K}$ (mpc)\\ [0.5ex]
\hline 
0.1& $5\times 10^6$ & $1.5\times 10^4$ & $1.5\times 10^{-3}$ & $0.30$\\ 
1.& $5\times 10^5$ & $570$ & $5.7\times 10^{-4}$ & $0.17$ \\ 
10.& $5\times 10^4$ & $23$  & $2.3\times 10^{-4}$ & $0.096$\\  
[1ex] 
\hline 
\end{tabular} 
\label{Table:Series1} 
\end{table}

\begin{figure}[h!]
\includegraphics[width=7.5cm]{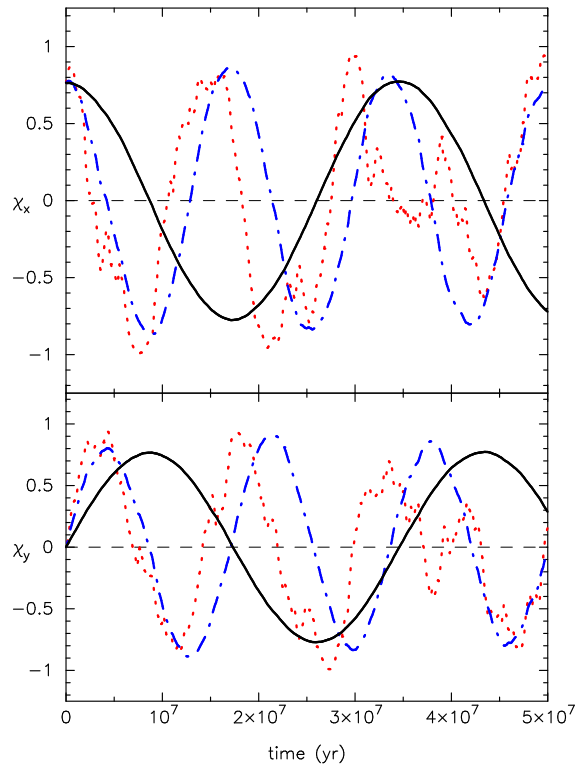}
\caption{\label{Figure:Series1} 
$x$ and $y$ components of the spin vector $\boldsymbol{\chi}\equiv c\boldsymbol{S}/G\mh^2$ for a set of integrations of fixed
$M_{0.1}=2\times 10^5\msun$ but different $N$ and $m_\star$:
$m_\star = 0.1\msun$ (thick/black), $m_\star=1\msun$ (dash-dotted/blue),
and $m_\star= 10\msun$ (dotted/red).
Other parameters of the models are given in Table~\ref{Table:Series1} and in 
the text.}
\end{figure}

Figure~\ref{Figure:Series1} shows the evolution of the \sbh\  spin vector in a set
of integrations with $f=0.1$ and with different $m_\star$; 
the number of stars was varied in
order to keep $M_{0.1}$ fixed at $2\times 10^5\msun$.
The correlation time $\tau$ was fixed at $10^5$ yr.
The number, $N_\mathrm{K}$, and total mass, $M_\mathrm{K}$, of
stars in the collisionless regime are listed in Table~\ref{Table:Series1}.
As $m_\star$ is increased (i.e. $N$ is decreased), the spin precessional
period drops, and the dependence of $\boldsymbol{S}$ on time
exhibits more stochasticity.
Both effects are consequences of the increasing number of stars
in the collisional regime.

Next we consider the nucleus of an intermediate-mass galaxy, 
$M_\mathrm{gal} \approx 10^{11}\msun$.
We set $\mh=10^8\msun$ and $m_\star=1\msun$.
The \sbh\ influence radius is $\rm\approx 35$ pc (Eq. \ref{Equation:rmvsmh}).
A typical value for $\gamma$ would be $\sim 2$
(Eq. \ref{Equation:GammaMean}), but given the large scatter in this
parameter, we consider a range of values.
Proceeding as before, we find from 
Eqs.~(\ref{Equation:DefineaSHigh}) and (\ref{Equation:DefineaKHighb}):
\begin{eqnarray}
a_\mathrm{S} &\approx& 0.21\, \mathrm{mpc} \ \ (\gamma=1) \nonumber \\
&\approx& 0.016\,\mathrm{pc} \ \ (\gamma=2)
\end{eqnarray}
and from Eq. (\ref{Equation:DefineaKHighb})
\begin{eqnarray}
a_\mathrm{K} &\approx& 0.16 \chi^{2/5}\, \mathrm{pc} \ \ (\gamma=1) \nonumber \\
&\approx& 0.042 \chi^{1/2}\, \mathrm{pc} \ \ (\gamma=2).
\end{eqnarray}
The  angular momentum of stars in the collisionless regime is given
by Eq. (\ref{Equation:ThetaKBright}) as
\begin{eqnarray}
\Theta_\mathrm{K} &\approx& 0.015 \left(\frac{f}{0.5}\right)\ \ \ \  \ \ (\gamma=1) \nonumber \\
 &\approx& 0.20 \chi^{-1/4} \left(\frac{f}{0.5}\right) \ \ (\gamma=2).\nonumber \\
\end{eqnarray}
$\Theta_\mathrm{K}$ approaches unity for sufficiently large values of $f$ and $\gamma$.
As in the previous example, almost all stars in the collisional regime have
$a\gap a_\mathrm{S}$, and for these stars Eq.~(\ref{Equation:DefinetMHigh}) gives
\begin{eqnarray}\label{Equation:tcohModel2}
t_\mathrm{coh} &\approx& 3.3\times 10^7 \left(\frac{a}{1\,\mathrm{pc}}\right)^{-1/2} \mathrm{yr} \ \ (\gamma=1) \nonumber \\
&\approx& 9.3\times 10^5 \left(\frac{a}{1\,\mathrm{pc}}\right)^{1/2} \mathrm{yr} \ \ (\gamma=2). 
\end{eqnarray}
Similarly
\begin{eqnarray}
T_\mathrm{RR,M} \approx 3\times 10^7 \left(\frac{m_\star}{\msun}\right)^{-1}
\left(\frac{a}{1\,\mathrm{mpc}}\right)^{3/2} \mathrm{yr}
\end{eqnarray}
and
\begin{eqnarray}
T_\mathrm{2dRR} &\approx& 1.2\times 10^7 
\left(\frac{m_\star}{M_\odot}\right)^{-1/2}
\left(\frac{a}{\mathrm{mpc}}\right)^{1/2} \mathrm{yr} \ \ (\gamma=1) \nonumber \\
&\approx& 6.2\times 10^4 
\left(\frac{m_\star}{M_\odot}\right)^{-1/2}
\left(\frac{a}{\mathrm{mpc}}\right) \mathrm{yr} \ \ (\gamma=2). \nonumber \\
\end{eqnarray}
The spin precessional period
due to torquing by stars in the collisional regime is 
\begin{eqnarray} \label{Equation:tspinModel2}
\frac{2\pi}{\omega_\mathrm{S}} &\approx&  7.2\times 10^{10} 
\left(\frac{f}{0.5}\right)^{-1} \chi^{1/5} \mathrm{yr} \ \ (\gamma=1)
\nonumber\\
&\approx&  5.0\times 10^8 
\left(\frac{f}{0.5}\right)^{-1} \chi^{3/4} \mathrm{yr} \ \ (\gamma=2).
\nonumber \\
\end{eqnarray}
Note the strong dependence of this time on $\gamma$ (Figure \ref{Figure:OmegaOU}).

\begin{figure}[h!]
\includegraphics[width=7.5cm]{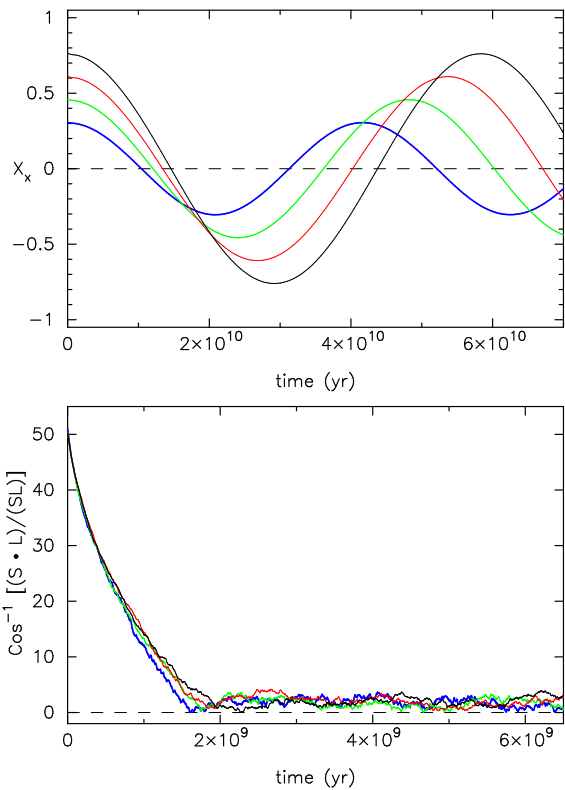}
\caption{\label{Figure:SpinThetag1} 
The upper panel shows evolution of the $x$ component of the \sbh\ spin in a set of integrations with $\mh=10^8\msun$, $\gamma=1$, $f=1/2$ and $\theta_0=50^\circ$.
The different curves correspond to different values of the dimensionless spin:
$\chi = 1.0$ (solid/black), $\chi=0.8$ (dash-dotted/red),
$\chi = 0.6$ (dashed/green), and $\chi=0.4$ (dotted/blue).
The lower panel shows the angle between $\boldsymbol{S}$ and $\boldsymbol{L}$,
the total angular momentum of stars in the ``collisionless'' regime
(i.e. near the \sbh).}
\end{figure}

\begin{figure}[h!]
\includegraphics[width=7.5cm]{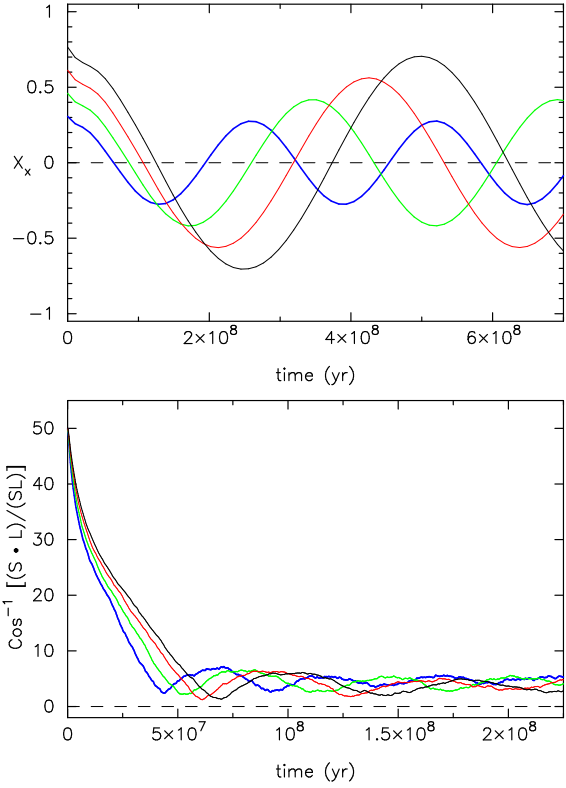}
\caption{\label{Figure:SpinThetag2} 
Like Figure \ref{Figure:SpinThetag1} but for $\gamma=2$.}
\end{figure}

Figures \ref{Figure:SpinThetag1} and \ref{Figure:SpinThetag2} show the
evolution of $\boldsymbol{S}$, and of the angle between $\boldsymbol{S}$ and 
$\boldsymbol{L}$, where $\boldsymbol{L}$ is the angular momentum of stars in the collisionless regime,
for models with $\gamma=\{1,2\}$, $f=0.5$, $\theta_0=50^\circ$,
and various values of $\chi$.
In both sets of model, the long-term precession rate of the \sbh\
depends modestly on $\chi$, and strongly on $\gamma$,
as expected from the relations (\ref{Equation:tspinModel2}).
There is an initial phase in which the stars in the collisionless regime
near the \sbh\ differentially precess about the nearly-fixed $\boldsymbol{S}$;
the length of this phase is $\sim 2$ Gyr for $\gamma=1$ and
$\sim 0.1$ Gyr for $\gamma=2$.
During this time, $\boldsymbol{S}$ reacts somewhat to the changing
$\boldsymbol{\omega}_\mathrm{K}$, before settling in to a more
regular precession (driven by $[\boldsymbol{\omega}_\mathrm{S}]_\mathrm{OU}$) at later times.

Particularly in the case $\gamma=2$, it is clear that the angle between $\boldsymbol{S}$
and $\boldsymbol{L}$ never reaches zero.
In this model, the time for stars within the
rotational influence sphere to differentially precess about $\boldsymbol{S}$ 
is $\sim 5 \times 10^7$ yr, only a few times smaller than the \sbh\ precessional period;
thus the differential precession can never quite ``catch up'' with the changing spin direction.
In the case $\gamma=1$, the ratio between these times is more than a factor 10
and the two vectors can nearly align.

\begin{figure}[h!]
\includegraphics[width=6.0cm,angle=-90.]{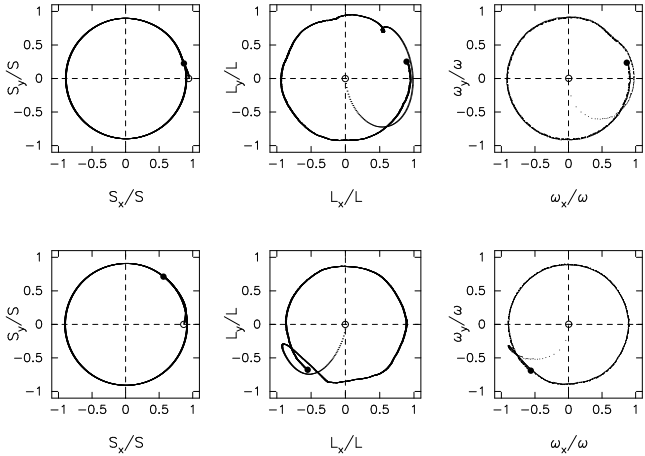}
\caption{\label{Figure:Two} 
Evolution of $\boldsymbol{S}$, $\boldsymbol{L}$ and $\boldsymbol{\omega}_\mathrm{S}$
in two models with $\chi=1$ and $\theta_0=70^\circ$ (upper) and 
$\theta_0=120^\circ$ (lower); other parameters are as in Figure \ref{Figure:SpinThetag1}.
The quantities $\boldsymbol{L}$ and $\boldsymbol{\omega}$ refer to the total angular
momentum, and the spin precessional vector, due to stars in the collisionless regime.
Open/filled circles are initial/final values; elapsed time is $0.5$ Gyr.}
\end{figure}

Figure \ref{Figure:Two} plots the evolution of $\boldsymbol{S}$, $\boldsymbol{L}$ and
$\boldsymbol{\omega}_\mathrm{S}$ for two models with $\gamma=2$, $\chi=1$ and
two values of $\theta_0=\{70^\circ, 120^\circ\}$; other parameters are as in 
Figure \ref{Figure:SpinThetag1}, and $\boldsymbol{L}$ and $\boldsymbol{\omega}_\mathrm{S}$ refer to stars in the collisionless regime only.
In these models, $\boldsymbol{L}$ is strongly misaligned with $\boldsymbol{S}$ initially,
and its direction evolves in a very complicated way at early times before nearly aligning with
$\boldsymbol{S}$.

\section{Discussion}
\label{Section:Discuss}

\subsection{Observations of nuclear rotation}
\label{Section:DiscussObserve}

\begin{figure}[h!]
\includegraphics[width=8.0cm,angle=0.]{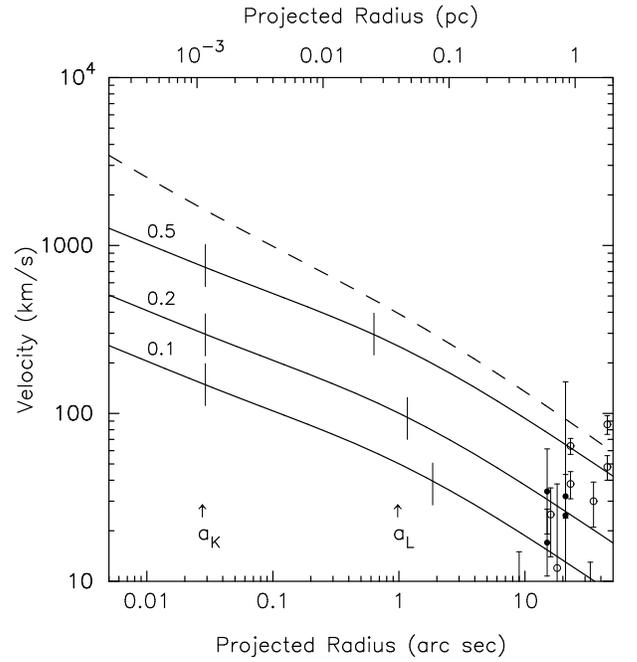}
\caption{\label{Figure:MWRotation} 
Solid lines are predicted, observed rotational velocity curves for a spherical stellar system
around a \sbh\ of mass $4\times 10^6\msun$ at a distance of 8 kpc, 
similar to the Milky Way nucleus.
Curves are labelled by the fraction $f$ of flipped orbits; $\gamma=1.5$ was assumed.
The nucleus is assumed to be observed from a point in a plane perpendicular to 
the axis of nuclear rotation, and the velocities have been averaged over a perpendicular distance of 1 arc second $\approx 0.04$ pc.
The dashed curve is the 1d velocity dispersion in the case that $f=0$.
Tick marks label $a_\mathrm{K}$, Eq. (\ref{Equation:DefineaKHighb}), 
and $a_\mathrm{L}$, Eq. (\ref{Equation:acritlow}), assuming $\chi=1$, $M_{0.1}=10^5\msun$, and $m_\star=1\msun$.
Circles with error bars are measured, binned, line-of-sight mean velocities of
stars from \cite{McGinn1989} (open) and \cite{Trippe2008} (filled).}
\end{figure}

Evolution of \sbh\ spins via the mechanism discussed here
is a strong function of the degree of rotation of the nucleus,
at distances $\rg\ll r \ll \rm$ from the \sbh, 
where  $\rg$ is the gravitational radius
(Eq. \ref{Equation:Definerg}) and $\rm$ the radius of influence
(Eq. \ref{Equation:Definerm}).
Observational constraints on the degree of rotation at such small radii tend
to be weak.
The nucleus of the Milky Way is the closest.
Figure \ref{Figure:MWRotation} plots line-of-sight, rotational velocity data for 
binned samples of stars at projected distances $\lap \rm\approx 2.5$ pc
$\approx 65^{\prime\prime}$ from Sgr A$^\star$ \cite{McGinn1989,Trippe2008}.
Also shown for comparison are  rotational velocity curves predicted by the
spherical models used here (\S\ref{Section:Models}); recall that
the degree of net rotation in those models is set by the parameter $f$, with
$f=1/2$ corresponding to maximal rotation.
The Milky Way data  are consistent with all values of $f$ but the  available data extend inward only to 
$\sim 0.2\rm$, well outside the region that would  contribute most of  the torque
to a spinning \sbh.

The complexity of the velocity data has led to suggestions \cite{Schoedel2009}
that the Milky Way nucleus consists of a superposition of different structures
with different axes of rotation.
While the stars contributing to the velocity data in Figure \ref{Figure:MWRotation}
are mostly old, 
two disklike structures of young stars -- the clockwise disk discussed above,
and another (the ``counter-clockwise disk'') \cite{Paumard2006}, both at
$\sim 0.1$ pc -- are known to rotate about axes that are separated by $\sim 60 ^\circ$ and both disks are inclined with respect to the large-scale symmetry plane of the Galactic disk.

The Local Group dwarf galaxy NGC 221 (M32), 
at a distance of $\sim 700$ kpc \cite{McConnachie2005}, also appears to contain
a \sbh, with mass that is poorly constrained but believed to be similar to that of the Milky Way \sbh\ \cite{Valluri2005}.
The line-of-sight mean stellar velocity in this galaxy is roughly constant with radius,
$v_\mathrm{rot}\approx 60 $ km s$^{-1}$ \cite{Joseph2001,Verolme2002},
inside a projected radius of $\sim \rm\approx 3$ pc,
compared with a line-of-sight velocity dispersion $\sigma\approx 75$ km s$^{-1}$,
suggesting $f\lap 0.5$.
The resolution in this case is $\sim 0.3$ pc $\sim 0.1\rm$.

Beyond the Local Group, massive galaxies are the best prospects for
spatially resolving a region smaller than $\rm$ due to the scaling of $\rm$
with galaxy mass.
A region  of size $0.1\rm$ in a galaxy at distance $D$ has  angular extent
\beq
\theta(0.1 \rm) \approx 0^{\prime\prime}.07 \left(\frac{\mh}{10^8\msun}\right)^\alpha
\left(\frac{D}{10\,\mathrm{Mpc}}\right)^{-1}
\eeq
where $\alpha\approx 0.56$ (Eq. \ref{Equation:rmvsmh}).
Observations rarely exceed the 
$\sim 0^{\prime\prime}.1$ resolution of STIS on the {\it Hubble Space Telescope}
and so little data are available on scales much less than $0.1\rm$ for galaxies
beyond the Local Group.
Nevertheless, a number of nearby galaxies exhibit strong nuclear rotation,
$v_\mathrm{rot}\sim\sigma$ on the smallest resolvable scales.
Some examples (NGC number, 
followed by the radius of the resolved region,
expressed  as a fraction of $\rm$) are:
NGC 3115 ($0.04$) \cite{Emsellem1999},
NGC 3377 ($0.20$) \cite{Copin2004},
NGC 3379 ($0.10$) \cite{Shapiro2006},
NGC 4342 ($0.12$) \cite{Cretton1999},
NGC 4258 ($0.17$) \cite{Siopis2009}.
Data like these are at least consistent with the presence of 
significant nuclear rotation on spatial scales $\ll\rm$
although of course they do not compel it.
As noted above (\S\ref{Section:Models}), in giant galaxies, the radius containing
an orbital angular momentum equal to $S$ is expected to be $\sim 0.01\rm-0.1\rm$ in
the case $v_\mathrm{rot}\sim\sigma$.

\subsection{Model constraints on the degree of nuclear rotation}
Given the weak observational constraints on rotation of
galactic nuclei, it is interesting to ask what  various models of
nuclear evolution predict.

Rotation arises naturally if stars formed  in a thin gaseous disk around
the \sbh\ \cite{LevinBeloborodov2003,NayakshinCuadra2005}, before later
being scattered (say) into more spheroidal structures.
Star formation requires that the gas disk be dense enough for its internal
gravity to overcome shearing and tidal stresses from the \sbh.
Steady-state accretion disk models \cite{CollinSouffrin1990} suggest a minumum radius for
star formation of \cite{KingPringle2007}
\begin{eqnarray}\label{Equation:rminaccretion}
r_\mathrm{min}&\approx& 10^{-2} \left(\frac{\alpha}{0.03}\right)^{14/27}
\left(\frac{\epsilon}{0.1}\right)^{8/27} \nonumber \\
&\times& \left(\frac{L}{0.1 L_\mathrm{E}}\right)^{-8/27}
\left(\frac{\mh}{10^8\msun}\right)^{1/27} \mathrm{pc}
\end{eqnarray}
where $\alpha$ is the standard viscosity parameter \cite{ShakuraSunyaev1973},
$L$ is the luminosity due to gas accretion onto the \sbh, 
$\epsilon$ is the accretion efficiency defined by $L=\epsilon\dot M c^2$, and
$L_\mathrm{E}\approx 1.4\times 10^{46}(\mh/10^8\msun)$ erg s$^{-1}$
is the Eddington luminosity.
The predicted dependence of $r_\mathrm{min}$ on $\mh$ is extremely weak.

An $r_\mathrm{min}$ of $10^{-2}$ pc is similar to the
inner radius of the young ``clockwise disk'' of stars at the Galactic center
\cite{Bartko2010,Nayakshin2006,Paumard2006}.
However there is currently no evidence of an accretion disk
\cite{Cuadra2003,NayakshinCuadraSunyaev2004} and the low luminosity of 
Sgr A$^\star$ places strict limits on its current rate of gas accretion \cite{Bower2003}.
Attempts to  explain the formation of the young stars usually 
invoke instead the recent infall and tidal shearing of a massive gas cloud.
Numerical simulations of this scenario
\cite{BonnellRice2008,HobbsNayakshin2009,Alig2011,Mapelli2012}
have confirmed that formation of a disk from which stars subsequently fragment
is possible if the initial conditions (cloud mass, density, temperature; 
orbital parameters) are correctly chosen.
Star formation in these models takes place as close as $\sim 0.01$ pc to the \sbh.
Given the small number of published simulations, and the fact that they were 
motivated by a desire to reproduce the known properties of the stellar disk,
is not clear whether different initial conditions might allow star formation much
farther in.

Figure \ref{Figure:Radii} suggests that for galaxies with $\mh\gap 10^7\msun$,
$a_\mathrm{K}\gap 10^{-2}$ pc.
For these galaxies, restricting the region of significant rotation to $r\gap 10^{-2}$ pc
would reduce somewhat the contribution to $\omega_\mathrm{S}$ from stars
in the ``collisionless'' regime but would not change the implied rate of steady \sbh\ 
precession due to stars beyond $a_\mathrm{K}$,  as given by Eq. (\ref{Equation:OmegaOU}).
In the case of low-mass galaxies, removing stars inside $\sim 10^{-2}$ pc would 
essentially turn off the collisionless contribution to $\dot{\boldsymbol S}$ and
increase the \sbh\ precessional period due to stars in  the collisional regime
by an approximate factor
$(a_\mathrm{K}/r_\mathrm{min})^{1/2-\gamma}$.

Another possible source of nuclear rotation is inspiral of a massive object,
which transfers its orbital angular momentum to the stars via dynamical friction
before being captured by the \sbh\ (say).
Assume that the inspiralling object has a mass $m_\bullet$, where
$m_\star\ll m_\bullet \ll \mh$.
Assuming a circular orbit, a decrease in orbital radius of $\Delta r$ implies
a transfer to the stars of angular momentum
\beq
\Delta L = \frac{m_\bullet}{2}\sqrt{\frac{G\mh}{r}} \Delta r.
\eeq
We want to compare this with the maximum, net angular momentum that could
be associated with the stars in a shell of thickness $\Delta r$:
\beq
\Delta L_\star = f \times r \times \sqrt{\frac{G\mh}{r}} \times 4\pi r^2\rho(r)\, \Delta r
\eeq
where $f\lap 1$ depends on the morphology of the nucleus and the 
distribution of stellar orbits.
Thus
\beq
\left|\frac{\Delta L}{\Delta L_\star}\right| = \frac{1}{8\pi f}\frac{m_\bullet}{\rho r^3}.
\eeq
In terms of the density model adopted here for low-mass galaxies, 
this can be expressed as
\beq
\left|\frac{dL}{dL_\star}\right| = \frac{1}{2(3-\gamma)f} \frac{m_\bullet}{M_{0.1}}
\left(\frac{r}{0.1\,\mathrm{pc}}\right)^{\gamma-3}.
\eeq
Since $\gamma < 3$, this result implies that the largest fractional increase
in orbital angular momentum occurs for stars nearest the \sbh.
The model must break down at radii where the enclosed stellar mass
is less than $\sim m_\bullet$, or
\beq
r \approx 0.1\,\mathrm{pc} \left(\frac{m_\bullet}{M_{0.1}}\right)^{1/(3-\gamma)}.
\eeq
For example, setting $m_\bullet=10^3\msun$  (an ``intermediate-mass black hole'');
$M_{0.1}=10^5\msun$; and $\gamma=2$, we find $r_\mathrm{min}\approx 1$ mpc.
At this radius, $|dL/dL_\star|$ is maximized and equal to $1/[2(3-\gamma)f]$
which can be of order unity.

\begin{figure}
\includegraphics[angle=-90]{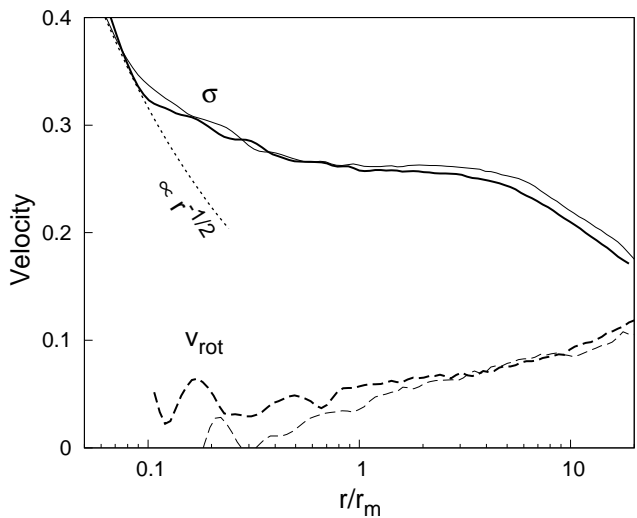}
\caption{\label{Figure:RotationInMergerSim} 
Rotation in the galaxy merger simulations of Gualandris \& Merritt (2012)
\cite{GualandrisMerritt2012}.
Solid lines are the one-dimensional velocity dispersion, dashed lines
are the rotation velocity about the $z$ axis (the orbit of merging galaxies was the $x-y$ plane).
Thick (thin) lines are for a merger on a circular (eccentric) orbit.
The galaxy mass ratio was 1:3 and each galaxy hosted a 
\sbh\ with $M_\bullet=0.005\,M_\mathrm{galaxy}$.
The models contained four different stellar masses representing an old stellar 
population; stars from all mass groups were weighted equally in constructing
this figure.
The dotted line shows a Keplerian rise in velocity dispersion and $\rm$ is
the radius containing a stellar mass equal to twice the combined mass
of the two \sbhs.
}
\end{figure}

Somewhat larger changes in nuclear structure and kinematics would result from the
dissipationaless (gas-free) merger of two galaxies containing comparably-massive 
\sbhs\ \cite{MilosavljevicMerritt2001}.
This is a likely model for the origin of the cores that are observed at the centers of galaxies
with $\mh\gap 10^{7.5}\msun$ \cite{Merritt2006}. 
Orbital motion of the two galaxies would imprint rotation on the stars in the
merged galaxy, but the binary \sbh\  also displaces a mass in stars
of order its own mass via the gravitational slingshot \cite{MikkolaValtonen1992};
the net rotation of the stars left behind depends in a complicated way on this process.
Figure~\ref{Figure:RotationInMergerSim} shows results extracted from 
perhaps the highest-resolution study to date of this interaction 
\cite{GualandrisMerritt2012}.
The figure shows the velocity dispersion 
and rotation velocity profiles of the merged galaxy at the time when two \sbhs\ coalesce.
Despite limited statistics at small radii, it is clear that such merger products 
may have a noticeable degree of rotation well within the radius of influence, 
corresponding roughly to $f\approx 0.1-0.2$.

Both infall of gas clouds and inspiral of massive compact objects could occur
episodically.
In the case of spiral galaxies,  the example
of the Milky Way with its young stellar disks suggests that accretion events
might occur roughly once per $0.1$ Gyr \cite{Alexander2005}.
In the case of massive galaxies, which tend to be gas-poor,
large-scale simulations of dark-matter clustering
suggest that the mean time between galaxy mergers  varies from $\sim 0.2$ Gyr at a redshift $z=10$ to $\sim 10$ Gyr at $z=1$ with a weak dependence on galaxy (i.e. dark
halo) mass \cite{Fakhouri2010}.
Assuming that all or most galaxies contain nuclear \sbhs, 
this would also be roughly the time between insertion of secondary
\sbhs\ into the nucleus \cite{BBR1980}.
These times are  comparable with the time scales for spin
precession derived here (e.g. Figure \ref{Figure:OmegaOU}), 
suggesting that the evolution of \sbh\
spins due to frame dragging may also be episodic in nature.

Both sorts of infall event are likely to occur from essentially 
random directions,  so that the increase over time of the net rotation of
the nucleus will have the form of a random walk.
Futhermore, both sorts of event can change the magnitude of $\boldsymbol{S}$:
if accretion of the gas by the \sbh\ occurs \cite{KingPringle2007};
or if the inspiralling body coalesces with the \sbh\ \cite{MerrittEkers2002}.


\subsection{Comparison with accretion disk torquing}
\label{Section:DiscussDisk}

Interaction of a spinning \sbh\ with a misaligned, gaseous accretion disk is driven
by the same frame-dragging torques modelled here.
The accretion disk problem has been extensively studied
\cite{BardeenPetterson1975,Rees1978,Sarazin1980,ScheuerFeiler1996, NatarajanPringle1998, KLOP2005}, 
in part because 
the radio jets that power the classic, double radio sources are 
believed to be launched perpendicularly to the inner accretion disk.
Both the long-term ($\sim 10^8$ yr) stability of jet directions in some 
active galaxies,
as well as the evidence for jet precession in others, is probably linked in fundamental
ways to \sbh\ - accretion disk interactions.

Here we sketch the points of similarity and difference between spin evolution driven by 
a misaligned accretion disk and by a rotating stellar nucleus.
We emphasize that the latter case is generic -- \sbhs\ appear {\it always} to be
embedded in stellar nuclei -- while nuclear activity, hence accretion disks,
exist in only a small subpopulation of galaxies.
The fact that accretion disk torques have received essentially all the attention until
now is probably a consequence of the easy observability of the jets.

Given an assumed structure for the disk (surface density and inclination as 
functions of radius), the instantaneous evolution equation
for $\boldsymbol{S}$ is essentially Eq.~(\ref{Equation:SpinEvol}),
after setting orbital eccentricities in that equation to zero and identifying $\boldsymbol{L}_j$
with the angular momentum of a discrete  element of gas.
Such models typically assume a disk that is thin and initially planar.
Differential precession then ensues near the \sbh; at radii $r\ll r_\mathrm{L}$ --
defined, as in Eq.~(\ref{Equation:acritlow}), as the radius containing an angular momentum
equal to $S$ -- the gas precession time is short compared with that of the \sbh.

The value of $r_\mathrm{L}$ can be computed given a model for the disk surface density.
For instance, the steady-state disk models referred to above \cite{CollinSouffrin1990} imply
\cite{KLOP2005}
\begin{eqnarray}\label{Equation:rLAGN}
r_\mathrm{L}&\approx& 0.052 \chi^{10/19} \left(\frac{\alpha}{0.03}\right)^{8/19}
\left(\frac{\epsilon}{0.3}\right)^{6/19} \nonumber \\
&\times&\left(\frac{\mh}{10^8\msun}\right)^{7/19}
\left(\frac{L}{0.1\, L_\mathrm{E}}\right)^{-6/19} \mathrm{pc}
\end{eqnarray}
where $\alpha$, $\epsilon$ and $L_\mathrm{E}$ are defined as in 
Eq. (\ref{Equation:rminaccretion}).
In active galaxies, all quantities in parentheses aside from the factor containing
$\mh$ are of order unity and so $10^{-2.5} \mathrm{pc}\lap r_\mathrm{L}\lap 10^{-0.5}$ pc.
This is somewhat smaller than the value of $a_\mathrm{L}$ as plotted in Figure
\ref{Figure:Radii}, at least in massive galaxies; in other words: 
accretion disks, when present, are likely to dominate the angular momentum 
distribution near the \sbh, justifying the neglect of stellar torques in these galaxies.
In smaller galaxies, Figure \ref{Figure:Radii} suggests that $a_\mathrm{L}\sim r_\mathrm{L}$.

Differential precession causes gas near the \sbh\ to attain 
a mean $\boldsymbol{L}$ that is aligned with $\boldsymbol{S}$, 
as in the stellar case, but gaseous
viscosity also ensures that the gas returns to a thin disk, coincident with
the \sbh\ equatorial plane (the ``Bardeen-Petterson effect'' \cite{BardeenPetterson1975}).
This thin, aligned disk extends outward, not to $r_\mathrm{L}$ (where the Lense-Thirring
time is likely to be very long anyway), but rather to the smaller radius $r_\mathrm{warp}$, the radius at which the disk plane transitions to its large-radius orientation.
The warp radius is determined by the condition 
that the time scale for angular momentum diffusion through the disk
is equal to the Lense-Thirring precession time: 
\begin{equation}
t_\mathrm{diff} \approx \frac{r_\mathrm{warp}^2}{\nu} \;\approx\;
\omega_\mathrm{LT}^{-1} = \frac{r_\mathrm{warp}^3 \, c^2}{2GS}.
\end{equation}
Here $\nu$ is the kinematic viscosity, which also determines the accretion rate.
The value of $r_\mathrm{warp}$ is strongly model-dependent and still rather uncertain; 
early estimates (e.g. \cite{BardeenPetterson1975,Sarazin1980}) set $r_\mathrm{warp}\approx r_\mathrm{L}$,  but more recent estimates (e.g. \cite{NatarajanPringle1998,Martin2007})
 find $r_\mathrm{warp}\approx (10^2-10^3)\times \rg\ll r_\mathrm{L}$.

Once alignment of the gas inside $r_\mathrm{warp}$ 
has occurred, precession of the \sbh\ is driven by gas at $r\gap r_\mathrm{warp}$.
An expression that is often given for the steady-state \sbh\ precession frequency
(e.g. \cite{Sarazin1980,NatarajanPringle1998}) is
\beq\label{Equation:omegawarp}
\omega \approx \frac{L(r<r_\mathrm{warp})}{S} \times \omega_\mathrm{LT}(r_\mathrm{warp})
\eeq
where $L(r<r_\mathrm{warp})$ is the angular momentum of disk gas inside
$r_\mathrm{warp}$.
Uncertainties about the value of $r_\mathrm{warp}$ translate via this expression
into uncertainties about the precession rate.
Equation (\ref{Equation:omegawarp}) is similar to Eq. (\ref{Equation:OmegaK}) 
for the mutual precession of a \sbh\ and a ring of matter, especially when it is recognized that $\boldsymbol{J}\approx \boldsymbol{S}$ in the accretion-disk case.
This is at first sight surprising, since Eq. (\ref{Equation:omegawarp}) appears to relate the
precession of the \sbh\ to the angular momentum of gas all of which, by assumption,
is fully aligned with the \sbh!
The justification (e.g. \cite{Sarazin1980}) consists of noting that $L\times\omega_\mathrm{LT}\propto L(r)/r^3$ is a steeply falling function of radius, 
hence only matter near the warp is relevant.
But this argument underscores the very approximate nature of Eq. (\ref{Equation:omegawarp}).

The warp radius plays approximately the same role as the radius $a_\mathrm{K}$
in the stellar case, Eq. (\ref{Equation:DefineaK}).
The \sbh\ precession frequency, Eq. (\ref{Equation:omegawarp}) in the gaseous case,
becomes Eq. (\ref{Equation:OmegaOU}) in the stellar case.
In the stellar case, $a_\mathrm{K}\ll a_\mathrm{L}$ (Figure \ref{Figure:Radii}), 
just as  $r_\mathrm{warp}\ll r_\mathrm{L}$ (at least if
the most recent estimates of $r_\mathrm{warp}$ are correct).

The continued deposition of matter from a fixed outer plane must ultimately
align $\boldsymbol{S}$ with the outer $\boldsymbol{L}$. 
In many models \cite{ScheuerFeiler1996,NatarajanPringle1998,LodatoPringle2006}, 
the time scale for this alignment is similar 
to the warp-driven precession time of the \sbh, i.e. the inverse
of Eq. (\ref{Equation:omegawarp}). 
Typical values quoted for 
$t_\mathrm{align}$ lie in the range $10^7-10^8$ years and it has been argued
that this alignment is responsible for the long-term ($10^8-10^9$ yr) stability
of jet directions in many active galaxies.
Interestingly, we  found that complete alignment was possible also in the
stellar-dynamical case  (Figure \ref{Figure:sphere40});  differential precession
is sufficient to achieve this, even in the  absence of viscosity.
However we argued that a more generic outcome in the stellar case is steady precession
of the \sbh, particularly when stellar interactions are allowed.


The evolution of $\boldsymbol{S}$ due to the {\it combined} influence of a
misaligned accretion disk and stars is beyond the scope of this paper, but we 
include a few speculative remarks \footnote{Torquing or heating of an accretion disk
by stars has been considered by a number of authors \cite{Ostriker1983,Vilkoviskij1983,BregmanAlexander2012}; 
the latter authors also considered 
reaction of \sbh\ spin to changes in the accretion disk.}
Feeding of active galaxies is probably episodic
\cite{Schoenmakers2000,Kharb2006}.
When much, but not all, of the infalling gas has been consumed, there may come
a time when the precession rates due to gas and stellar torquing are comparable.
If the \sbh\ is still active at this time, accretion-disk-related jets should begin to precess
roughly in the manner discussed here, even if the \sbh\ had previously reached a 
steady-state alignment with the gas.
Prolonged, steady precession of radio sources might be explained 
in this way \cite{Gower1982,LuZhou2005}.
After the gas has been fully consumed, 
the \sbh\ spin can continue to evolve in response to the stars.
If the gas has been accreted all the way to the event horizon, 
both the magnitude and direction of $\boldsymbol{S}$ will have been changed
by the gas.

\subsection{Slowly-rotating nuclei}

Even in a nucleus with negligible net rotation, there will still
be a nonzero  torque on the \sbh\ due to imperfect cancellation of the 
$\boldsymbol{L}_j$ from the finite number of stars.
This is obvious, for instance, from Figure \ref{Figure:omega42};
the components of $\boldsymbol{\omega}_\mathrm{S}$ perpendicular
to the mean rotation axis of the cluster are zero on average but 
fluctuate as orbits change their $\boldsymbol{L}_j$ due to encounters.
 Eq. (\ref{Equation:sigmaest}) is an estimate of the size of those fluctuations
and can equally well be interpreted as the expected value of $\omega_\mathrm{S}$ in a nonrotating, isotropic cluster with known $N(a)$.

It is interesting to ask how large the steady rotation of a nucleus needs to be
if the net torque exerted on the \sbh\ is to exceed this (fluctuating) value.
We estimate the torque in a {\it non}rotating nucleus by setting $r_\mathrm{p,min}=a_\mathrm{K}$ in Eq. (\ref{Equation:sigmaest}); in other words,
we conservatively ignore the torque from stars within the sphere of rotational influence
given that they may have differentially precessed about $\boldsymbol{S}$.
The result is
\beq
\sigma^2 \approx \frac43\frac{(3-\gamma)\Gamma(\gamma+1)}{\Gamma(\gamma+4)}
\frac{G^3\mh m_\star}{c^4}\frac{M_0}{r_0^5}
\left(\frac{a_\mathrm{K}}{r_0}\right)^{-(2+\gamma)}
\eeq
where $\{M_0,r_0\}=\{M_{0.1}, 0.1\,\mathrm{pc}\}$ in low-mass galaxies and
$\{M_0,r_0\}=\{2\mh,\rm\}$ in high-mass galaxies.
Comparing $\sigma$ to $\omega_\mathrm{S}$ as given by Eq. (\ref{Equation:OmegaOU}),
we find for the critical degree of rotation
\begin{eqnarray}
f &\approx& J(\gamma) \sqrt{\frac{m_\star}{M_0}}
\left(\frac{a_\mathrm{K}}{r_0}\right)^{(\gamma-3)/2}, \\
J(\gamma) &=& \frac14 \sqrt{\frac{3}{\pi}\frac{\Gamma(\gamma+1)}{\Gamma(\gamma+4)}}
\frac{\Gamma(\gamma-1/2)}{\Gamma(\gamma+1)}
\frac{2^\gamma(2\gamma-1)^2}{(3-\gamma)(6-\gamma)} .\nonumber
\end{eqnarray}
For instance, in a low-mass galaxy with $\gamma=2$,
\beq
f \approx 8\times 10^{-3} \chi^{-1/4} \left(\frac{\mh}{10^6\msun}\right)^{-5/8}
\left(\frac{M_{0.1}}{10^5 m_\star}\right)^{-3/8}. 
\eeq
For this value of $f$, the instantaneous time over which $\boldsymbol{S}$ changes is
\beq\label{Equation:tcrit}
\frac{2\pi}{\omega_\mathrm{S}} \approx 8\times 10^8\, \chi \left(\frac{\mh}{10^6\msun}\right)^8
\left(\frac{M_{0.1}}{10^5\msun}\right)^{-1} \mathrm{yr}.
\eeq
We emphasize that this is not a precessional {\it period} since, by assumption,
finite-$N$ effects dominate and the axis about which $\boldsymbol{S}$ is 
precessing will itself change, in a time of order $T_\mathrm{2dRR}$.

\begin{figure}
\includegraphics[width=8.cm]{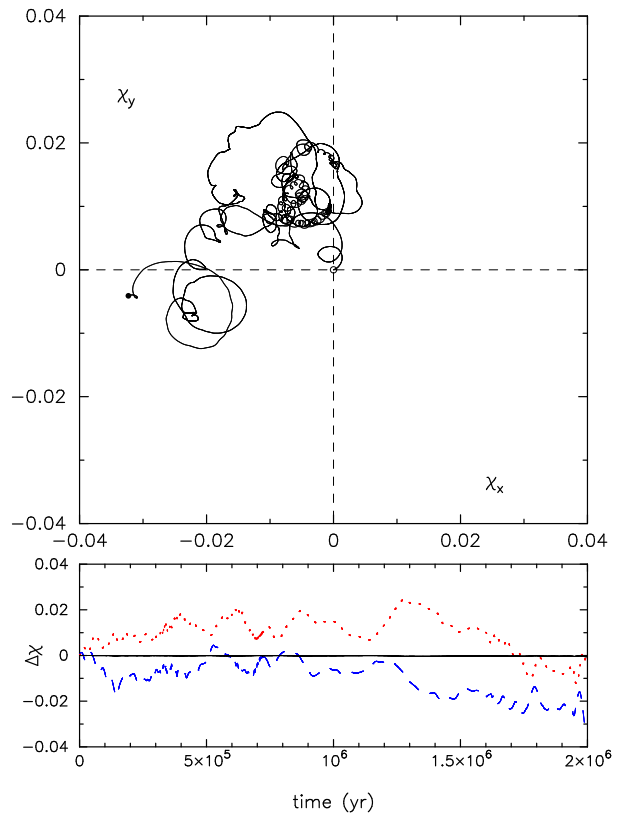}
\caption{\label{Figure:NoRotate} Evolution of the dimensionless spin
in a direct $N$-body integration \cite{MAMW2011}.
The top panel shows the $x$ and $y$ components of $\boldsymbol{\chi}$
which is initially parallel to the $z$-axis.
Open and filled circles show initial and final orientations.
In the bottom panel, the dashed (blue) curve is $\chi_x$, the
dotted (red) curve is $\chi_y$, and the solid (black) curve is $\chi_z$.
}
\end{figure}

When observing changes in $\boldsymbol{S}$ over very short time scales,
e.g. human lifetimes, there will also be a time-dependent contribution due to the motion of
stars along their (unperturbed) orbits.
This contribution has been ignored up till now due to the orbit-averaging
of Eqs. (\ref{Equation:SpinEvol})-(\ref{Equation:Ljdot}).

Figure~\ref{Figure:NoRotate} illustrates the complexity of the evolution of 
$\boldsymbol{S}$ in the case that all the torque on the \sbh\ is due to
these finite-$N$ effects.
The figure is based on a direct $N$-body integration of a cluster of $50$ 
``stars,'' of mass $50\msun$ each,
around a \sbh\ of mass $10^6\msun$ and $\chi=1$.
Additional details about the initial models are given in Merritt et al. (2011)
\cite{MAMW2011}.
Over the $\sim 2$ Myr time span of the integration, the \sbh\ spin axis wobbles by
about one degree.

\subsection{Experimental determination of black hole spins}
Several authors \cite{LevinBeloborodov2003,WIll2008,Eisenhauer2011} 
have suggested that it may be possible to infer the magnitude and direction
of \sbh\ spins from the precession of the angular momentum vectors of individual stars.
Only stars with semimajor axes $a\lap a_\mathrm{K}$ are suitable for this purpose, otherwise the changes of ${\boldsymbol L}_j$ 
due to collisional effects will supersede changes due to frame dragging
\cite{MAMW2010,SadeghianWill2011}. 
Since for most galaxies $a_\mathrm{K} \ll a_\mathrm{L}$ (Figure \ref{Figure:Radii}), 
this also means that the Lense-Thirring precession times 
for these stars are much shorter than \sbh\ spin precession times, or 
$\omega_j \gg \omega_S$.

Assuming that $\boldsymbol{S}$ precesses steadily about the (fixed) axis 
$\boldsymbol{\omega}_\mathrm{S}$, it is convenient to consider the evolution of 
$\boldsymbol{L}_j$ in the reference frame which rotates with the precession 
frequency $\boldsymbol{\omega}_\mathrm{S}$ about the axis of SBH precession, so that $\boldsymbol{S}$ is stationary. 
In this rotating frame, the equation of motion for the orbital angular momenta
(\ref{Equation:Ljdota}) reads
\begin{equation}
\dot{\boldsymbol L}_j^\prime = \boldsymbol{\omega}_j^\prime\times\boldsymbol{L}_j^\prime, \;\;\quad
\boldsymbol{\omega}_j^\prime \equiv \boldsymbol{\omega}_j - \boldsymbol{\omega}_S
\end{equation}
where ${\boldsymbol L}_j^\prime$ is the angular momentum in the rotating frame.
In other words: the Lense-Thirring precession is occuring about an axis that does not 
coincide with the instantaneous direction of $\boldsymbol{S}$. 
However, the first derivative 
of ${\boldsymbol L}_j$ in the inertial frame coincides with the value obtained without 
taking into account \sbh\ precession, since it is determined by the instantaneous value and 
direction of ${\boldsymbol S}$. It is only for the second derivative of ${\boldsymbol L}_j$ 
that the difference starts to matter. 
Given that $|\boldsymbol{\omega}_j^\prime-\boldsymbol{\omega}_j| \ll |\boldsymbol{\omega}_j|$ 
because we are considering stars that themselves precess much faster than the \sbh, 
it seems unlikely that these effects may be detectable in the near future.

In a nucleus that is sufficiently old, differential precession of stars 
with $a<a_\mathrm{K}$ will have caused their angular momentum vectors to distribute themselves uniformly about $\boldsymbol{S}$, as shown above in several numerical examples.
This suggests a way of measuring the instantaneous direction of $\boldsymbol{S}$ 
via the mean direction of the $\boldsymbol{L}_j$ within $a_\mathrm{K}$.
In the Galactic center, $a_\mathrm{K}\approx 10^{-3}$ pc, and there are as yet no
stars with determined orbits in this region.
However orbital periods for stars with $a=1$ mpc are about one year, 
and it is possible that determination of the orbital elements of a few such stars might be feasible over a shorter time interval than  is required for measuring  changes in the $\boldsymbol{L}_j$ due to frame dragging.

\section{Conclusions}
\label{Section:Conclusions}

\noindent 1. In a galactic nucleus containing a spinning supermassive black hole (\sbh), frame dragging results in mutual torques between the stellar orbits and the \sbh.
The result is precession of both the \sbh\ spin, $\boldsymbol{S}$, 
and the angular momentum vectors, $\boldsymbol{L}_j$, of the individual stellar orbits,
with $\boldsymbol{S}+\boldsymbol{L}_\mathrm{tot} =
\boldsymbol{S} + \sum_j\boldsymbol{L}_j$ conserved. 
For stars at a single distance from the \sbh,
the controlling parameter is the ratio between $S$ and $L_\mathrm{tot}$.
If $S\gg L_\mathrm{tot}$, stellar orbits precess about the nearly fixed $\boldsymbol{S}$ 
with the Lense-Thirring period; while 
if $L_\mathrm{tot}\gg S$, $\boldsymbol{S}$  precesses about the nearly fixed
$\boldsymbol{L}_\mathrm{tot}$ with a period that is shorter by a factor
$S/L_\mathrm{tot}$.
The inner parsec of the Milky Way is known to contain stellar subsystems 
having $L_\mathrm{tot}\approx S$.

\smallskip

\noindent 2. Ignoring interactions between the stars, solutions of the coupled
equations for $\dot{\boldsymbol S}$ and $\dot{\boldsymbol L}_{j=1,\ldots,N}$
in spherical nuclei 
reveal two evolutionary modes in the case that $L_\mathrm{tot}>S$: continued
precession of $\boldsymbol{S}$ about $\boldsymbol{L}_\mathrm{tot}$; or
damped precession, in which $\boldsymbol{S}$ and $\boldsymbol{L}_\mathrm{tot}$ 
come into nearly complete alignment after one precessional period of the \sbh.
Even in the first  mode, differential precession of orbits 
near the \sbh\ causes their net angular momentum to align with $\boldsymbol{S}$,
reducing the torque that they exert on the \sbh.
Subsequent precession of the \sbh\ is  driven by torques from stars at $r\gap r_\mathrm{L}$,
where $r_\mathrm{L}$ is the radius enclosing a net angular momentum equal to $S$.

\smallskip

\noindent 3. Newtonian interactions between stars
can change their orbital angular momenta in a time shorter than Lense-Thirring
precessional times.
We define the ``radius of rotational influence,'' $a_\mathrm{K}$, 
around a Kerr \sbh\ as the radius inside of which torques due to  frame dragging 
act more quickly than torques from the other stars.
Typical values for this radius are $\sim 10^{-3}$ parsecs in dense nuclei like that
of the Milky Way, increasing to $\sim 10^0-10^1$ parsecs in nuclei containing the most
massive \sbhs.
The angular momentum associated with stars in this ``collisionless'' region 
near the \sbh\ is likely to
be much smaller than $S$ in nuclei of the smallest galaxies 
but may be comparable to $S$ in massive galaxies.

\smallskip
\noindent 4. 
Interaction between stars at $r>a_\mathrm{K}$ leaves the total angular momentum
of these stars unchanged, but results in random fluctuations of the individual $\boldsymbol{L}_j$ and hence in the torque which they exert on the \sbh.
We develop a stochastic model, based on the Ornstein-Uhlenbeck equation,
for the torque exerted by these stars and verify it by comparison with 
high-accuracy $N$-body simulations.
We argue that $d\boldsymbol{S}/dt$ can be approximated as the sum of two
terms: deterministic torques exerted by stars inside $a_\mathrm{K}$,
whose angular momenta evolve solely in response to frame-dragging; 
and a stochastically-fluctuating torque due to stars outside $a_\mathrm{K}$.

\smallskip
\noindent 5. Examples of stochastic evolution of $\boldsymbol{S}$ are presented
for various nuclear models.
Typical evolution consists of sustained precession, with periods that 
are highly dependent on nuclear parameters, but which are expected
to increase with increasing $\mh$: 
likely periods are $\sim 10^7-10^8$ yr for low-mass \sbhs\ in dense
nuclei, $\sim 10^{8}-10^{10}$ yr for \sbh\ with masses $\sim 10^8\msun$, and
$\sim 10^{10}-10^{11}$ yr for the most massive \sbhs.

\begin{acknowledgments}
DM was supported in part by the National Science Foundation under  
grant no. 08-21141 and by the National Aeronautics and 
Space Administration under grant no. NNX-07AH15G.
S. Trippe kindly provided data used in Figure  \ref{Figure:MWRotation} and
A. Gualandris assisted with Figure \ref{Figure:RotationInMergerSim}.
We thank T. Alexander, E. Blackman, P. Kharb, A. King, A Robinson, and C. Will for helpful discussions.
\end{acknowledgments}

\bigskip
\appendix
\begin{widetext}
\section{Time scale for change in orbital eccentricity}
\label{Appendix:A}

Here we present approximate expressions \cite{DEGN} for the time scales associated with changes in orbital eccentricity due to resonant relaxation \cite{RauchTremaine1996} and evaluate them 
for the power-law density model used in the text.

Define the ``apsidal coherence time'' $t_\mathrm{coh}$ as the shorter of the
two  precession times
$t_\mathrm{S}$ and $t_\mathrm{M}$ defined in \S\ref{Section:Encounters},
each evaluated at typical values of $e$; say, $e\approx 1/2$.
Comparison of Eqs. (\ref{Equation:DefinetM}) and (\ref{Equation:T2dRR})
shows that $t_\mathrm{M} \approx T_\mathrm{2dRR}/\sqrt{N}$
where $N$ is the number of stars at $r<a$: apsidal
precession due to the spherically-distributed mass 
acts more rapidly than $\sqrt{N}$ torques at changing orbital orientations.
For elapsed times short compared with $t_\mathrm{coh}$,  the  torque
due to all the local stars is therefore nearly constant, and the angular momentum of a typical
star responds by changing approximately linearly with time.
In this ``coherent resonant relaxation'' regime, all the components of
$\boldsymbol{L}_j$, i.e. both the orientation angles and the eccentricity $e_j$, 
change with characteristic time $T_\mathrm{RR,coh}$ given by
\begin{equation}\label{Equation:T2dRRAgain}
T_\mathrm{RR,coh} \approx \frac{P}{2\pi} \frac{\mh}{m_\star}\frac{1}{\sqrt{N}}
\approx 4.7\times 10^4 \left(\frac{a}{\mathrm{mpc}}\right)^{3/2} 
\left(\frac{\mh}{10^6\msun}\right)^{-1/2} 
\left(\frac{\mh}{10^6 m_\star}\right)
\left(\frac{N}{10^2}\right)^{-1/2} \mathrm{yr}. 
\end{equation}
This is the same expression as Eq.~(\ref{Equation:T2dRR}) for $T_\mathrm{2dRR}$,
reflecting the fact that  in the coherent regime, 
both the direction and the magnitude of $\boldsymbol{L}$ change on roughly the same time scale.

For time intervals longer than $t_\mathrm{coh}$,
the direction of the net field-star torque changes, and evolution of
the $\boldsymbol{L}_j$ in response to the torques
is better described as a random walk.
The time scale associated with this ``incoherent resonant relaxation'' 
is
\beq\label{Equation:DefineTRR}
T_\mathrm{RR} \approx \left(\frac{L_c}{\Delta L_\mathrm{coh}}\right)^2 t_\mathrm{coh}
\eeq
where $L_c=\sqrt{G\mh a}$ is the angular momentum of a circular
orbit of semimajor axis $a$, and $\Delta L_\mathrm{coh}$ is the change in $L$
during $\Delta t = t_\mathrm{coh}$.
Setting $t_\mathrm{coh}=t_\mathrm{M}$ (i.e. $a>a_\mathrm{S}$), this becomes
\begin{equation}\label{Equation:TRRM}
T_\mathrm{RR,M}(a) = C_\mathrm{M}\frac{\mh}{m_\star} P(a) 
\approx 3\times 10^9 C_\mathrm{M} \left(\frac{\mh}{10^6\msun}\right)^{1/2}
\left(\frac{m_\star}{1 \msun}\right)^{-1} \left(\frac{a}{0.1 \mathrm{pc}}\right)^{3/2} \mathrm{yr}
\end{equation}
with $C_\mathrm{M}$ a constant of order unity, 
while if $t_\mathrm{coh}=t_\mathrm{S}$ ($a<a_\mathrm{S}$),
\begin{equation}\label{Equation:TRRS}
T_\mathrm{RR,S}(a) = C_\mathrm{S}\frac{\rg}{a}\left(\frac{\mh}{m_\star}\right)^2
\frac{P(a)}{N(a)} 
\approx 1.5\times 10^5 C_\mathrm{S} \left(\frac{\mh}{10^6\msun}\right)^{5/2}
\left(\frac{m_\star}{1 \msun}\right)^{-1} 
\left(\frac{M_{0.1}}{10^4\msun}\right)^{-1}
\left(\frac{a}{0.1 \mathrm{pc}}\right)^{\gamma-5/2} \mathrm{yr}
\end{equation}
with $C_\mathrm{S}$ again of order unity.
Eqs. (\ref{Equation:TRRM}) and (\ref{Equation:TRRS}) are the appropriate time
scales to associate with changes in orbital eccentricity in the incoherent regime.

In the case of 2d resonant relaxation, the relevant coherence time is that associated
with changes of the orbital planes, i.e. $T_\mathrm{2dRR}$.
Since $\Delta L_\mathrm{coh}\approx L_c$ in this case,
Eq.~(\ref{Equation:DefineTRR}) implies that 
the coherent and incoherent relaxation times are approximately the same:
no new time scale arises in the incoherent regime for 2d resonant relaxation.

Comparing the incoherent relaxation times associated with changes in the orientation
and magnitude of $\boldsymbol{L}$, respectively, we find
\begin{subequations}
\begin{eqnarray}
\frac{T_\mathrm{2dRR}}{T_\mathrm{RR,M}} &\approx& \frac{1}{\sqrt{N}}, \\
\frac{T_\mathrm{2dRR}}{T_\mathrm{RR,S}} &\approx& \frac{m_\star\sqrt{N}}{\mh}\frac{a}{\rg}.
\end{eqnarray}
\end{subequations}
The first of these ratios is manifestly smaller than unity at all radii.
The second is only relevant at $a\lap a_\mathrm{S}$, i.e. for
\beq
\frac{a}{\rg}\sqrt{N} \lap \frac{\mh}{m_\star}\frac{1}{\sqrt{N}}
\eeq
which  implies
\beq
\frac{T_\mathrm{2dRR}}{T_\mathrm{RR,S}} \lap \frac{1}{\sqrt{N}},
\eeq
again less than unity.
On the basis of these inequalities, it is reasonable to equate the correlation
time associated with fluctuations in $\boldsymbol{\omega}_\mathrm{S}$ with
the shortest of the time scales, $T_\mathrm{2dRR}$, as was done in \S\ref{Section:EncountersOU}.

\end{widetext}

\bibliography{biblio}
\end{document}